\documentclass[twocolumn]{aastex63}
\usepackage{epsf}
\usepackage{graphicx}
\usepackage{amsmath}
\usepackage{amssymb}
\usepackage{gensymb}
\usepackage{chngcntr}
\usepackage{longtable}
\usepackage{latexsym}
\received{}
\revised{}
\accepted{}
\shorttitle{Two Orion Cores in Deuterated Molecular Lines}
\shortauthors{Tatematsu et al.}

\begin{document}

\title{ALMA ACA and Nobeyama observations of two Orion cores in deuterated molecular lines}

\correspondingauthor{Ken'ichi Tatematsu}
\email{k.tatematsu@nao.ac.jp}

\author[0000-0002-8149-8546]{Ken'ichi Tatematsu}
\affil{Nobeyama Radio Observatory, National Astronomical Observatory of Japan, 
National Institutes of Natural Sciences, 
462-2 Nobeyama, Minamimaki, Minamisaku, Nagano 384-1305, Japan}
\affiliation{Department of Astronomical Science,
SOKENDAI (The Graduate University for Advanced Studies),
2-21-1 Osawa, Mitaka, Tokyo 181-8588, Japan}

\author[0000-0002-5286-2564]{Tie Liu}
\affiliation{Shanghai Astronomical Observatory, Chinese Academy of Sciences, 80 Nandan Road, Shanghai 200030, P. R. China}
\affiliation{Korea Astronomy and Space Science Institute,
Daedeokdaero 776, Yuseong, Daejeon 305-348, South
Korea}
\affiliation{East Asian Observatory, 660 N. A'ohoku Place, Hilo, HI 96720}

\author{Gwanjeong Kim}
\affil{Nobeyama Radio Observatory, National Astronomical Observatory of Japan, 
National Institutes of Natural Sciences, 
462-2 Nobeyama, Minamimaki, Minamisaku, Nagano 384-1305, Japan}

\author{Hee-Weon Yi}
\affiliation{School of Space Research, Kyung Hee University, Seocheon-Dong, Giheung-Gu, Yongin-Si, Gyeonggi-Do, 446-701, South Korea}

\author[0000-0003-3119-2087]{Jeong-Eun Lee}
\affiliation{School of Space Research, Kyung Hee University, Seocheon-Dong, Giheung-Gu, Yongin-Si, Gyeonggi-Do, 446-701, South Korea}

\author{Naomi Hirano}
\affiliation{Academia Sinica Institute of Astronomy and Astrophysics, 11F of Astronomy-Mathematics Building, AS/NTU. No.1, Sec. 4, Roosevelt Rd, Taipei 10617, Taiwan, R.O.C.}

\author[0000-0003-4603-7119]{Sheng-Yuan Liu}
\affiliation{Academia Sinica Institute of Astronomy and Astrophysics, 11F of Astronomy-Mathematics Building, AS/NTU. No.1, Sec. 4, Roosevelt Rd, Taipei 10617, Taiwan, R.O.C.}

\author{Satoshi Ohashi}
\affiliation{The Institute of Physical and Chemical Research (RIKEN), 2-1, Hirosawa, Wako-shi, Saitama 351-0198, Japan}

\author[0000-0002-7125-7685]{Patricio Sanhueza}
\affiliation{National Astronomical Observatory of Japan,
National Institutes of Natural Sciences,
2-21-1 Osawa, Mitaka, Tokyo 181-8588, Japan}
\affiliation{Department of Astronomical Science,
SOKENDAI (The Graduate University for Advanced Studies),
2-21-1 Osawa, Mitaka, Tokyo 181-8588, Japan}

\author{James Di Francesco}
\affiliation{Herzberg Astronomy \& Astrophysics, National Research Council of Canada, 5071 West Saanich Rd, Victoria, BC V9E 2E7, Canada}
\affiliation{Department of Physics and Astronomy, University of Victoria, Victoria, BC V8W 2Y2, Canada}

\author[0000-0001-5175-1777]{Neal J. Evans II}
\affiliation{Department of Astronomy, The University of Texas at Austin, 2515 Speedway, Stop C1400, Austin, TX 78712$-$1205, USA}

\author[0000-0001-8509-1818]{Gary A. Fuller}
\affiliation{Jodrell Bank Centre for Astrophysics, School of Physics and Astronomy, University of Manchester, Oxford Road, Manchester, M13 9PL, UK}

\author[0000-0003-2610-6367]{Ryo Kandori}
\affiliation{National Astronomical Observatory of Japan,
National Institutes of Natural Sciences,
2-21-1 Osawa, Mitaka, Tokyo 181-8588, Japan}

\author{Minho Choi}
\affiliation{Korea Astronomy and Space Science Institute,
Daedeokdaero 776, Yuseong, Daejeon 305-348, South
Korea}

\author{Miju Kang}
\affiliation{Korea Astronomy and Space Science Institute,
Daedeokdaero 776, Yuseong, Daejeon 305-348, South
Korea}

\author[0000-0002-4707-8409]{Siyi Feng}
\affiliation{The Kavli Institute for Astronomy and Astrophysics, Peking University,
5 Yiheyuan Road, Haidian District, Beijing 100871, P. R. China}
\affiliation{National Astronomical Observatory of Japan,
National Institutes of Natural Sciences,
2-21-1 Osawa, Mitaka, Tokyo 181-8588, Japan}

\author[0000-0003-1659-095X]{Tomoya Hirota}
\affiliation{National Astronomical Observatory of Japan,
National Institutes of Natural Sciences,
2-21-1 Osawa, Mitaka, Tokyo 181-8588, Japan}
\affiliation{Department of Astronomical Science,
SOKENDAI (The Graduate University for Advanced Studies),
2-21-1 Osawa, Mitaka, Tokyo 181-8588, Japan}

\author[0000-0003-4521-7492]{Takeshi Sakai}
\affiliation{Graduate School of Informatics and Engineering, The University of Electro-Communications, Chofu, Tokyo 182-8585, Japan}

\author[0000-0003-2619-9305]{Xing Lu}
\affiliation{National Astronomical Observatory of Japan,
National Institutes of Natural Sciences,
2-21-1 Osawa, Mitaka, Tokyo 181-8588, Japan}

\author{Quang Nguy$\tilde{\hat{e}}$n Lu'o'ng}
\affiliation{McMaster University, 1 James St N, Hamilton, ON, L8P 1A2, Canada}
\affiliation{Korea Astronomy and Space Science Institute,
Daedeokdaero 776, Yuseong, Daejeon 305-348, South
Korea}
\affiliation{National Astronomical Observatory of Japan,
National Institutes of Natural Sciences,
2-21-1 Osawa, Mitaka, Tokyo 181-8588, Japan}
\affiliation{IBM, Canada}

\author{Mark A. Thompson}
\affiliation{Centre for Astrophysics Research, Science \& Technology Research Institute, University of Hertfordshire, Hatfield, AL10 9AB, UK}

\author[0000-0002-5076-7520]{Yuefang Wu}
\affiliation{Department of Astronomy, Peking University, 100871, Beijing, China}

\author[0000-0003-3010-7661]{Di Li}
\affiliation{National Astronomical Observatories, Chinese Academy of Sciences, Beijing, 100012, China}

\author[0000-0003-2412-7092]{Kee-Tae Kim}
\affiliation{Korea Astronomy and Space Science Institute,
Daedeokdaero 776, Yuseong, Daejeon 305-348, South
Korea}

\author[0000-0002-7237-3856]{Ke Wang}
\affiliation{The Kavli Institute for Astronomy and Astrophysics, Peking University,
5 Yiheyuan Road, Haidian District, Beijing 100871, P. R. China}
\affiliation{European Southern Observatory, Karl-Schwarzschild-Str. 2 85748 Garching bei M\"{u}nchen, Germany}

\author{Isabelle Ristorcelli}
\affiliation{RAP, CNRS (UMR5277), Universit\'e Paul Sabatier, 9 avenue du Colonel Roche, BP 44346, 31028, Toulouse Cedex 4, France}

\author{Mika Juvela}
\affiliation{Department of Physics, P.O. Box 64, FI-00014, University of Helsinki, Finland}

\author[0000-0002-5310-4212]{L. Viktor T\'oth}
\affiliation{Department of Astronomy, E\"otv\"os Lor\'and Unviersity, P\'azm\'any P\'eter s\'et\'any 1/A, H-1117 Budapest, Hungary}





\begin{abstract}
We mapped two molecular cloud cores in the Orion A cloud with the ALMA ACA 7-m Array and with the Nobeyama 45-m radio telescope.  
These cores 
have bright N$_2$D$^+$ emission in single-pointing observations with the Nobeyama 45-m radio telescope,
have relatively high deuterium fraction,
and are thought to be close to the onset of star formation.
One is a star-forming core, and the other is starless.
These cores are located along filaments observed in N$_2$H$^+$, and 
show narrow linewidths of 0.41 km s$^{-1}$ and 0.45 km s$^{-1}$ in N$_2$D$^+$, respectively,
with the Nobeyama 45-m telescope.
Both cores were detected with the ALMA ACA 7m Array in the continuum and molecular lines at Band 6.
The starless core G211 shows clumpy structure with several sub-cores,
which in turn show chemical differences.
Also, the sub-cores in G211 have internal motions that are almost purely thermal. 
The starless sub-core G211D, in particular, shows a hint of the inverse P Cygni profile, suggesting infall motion.
The star-forming core G210 shows an interesting spatial feature of
two N$_2$D$^+$ peaks of similar intensity and radial velocity
located symmetrically with respect to the single dust continuum peak.
One interpretation is that the two N$_2$D$^+$ peaks
represent an edge-on pseudo-disk.
The CO outflow lobes, however, are not directed perpendicular to the line connecting both N$_2$D$^+$ peaks.
\end{abstract}

\keywords{ISM: clouds
---ISM: molecules
---ISM: structure---stars: formation}


\section{Introduction} \label{sec:intro}

Understanding initial conditions is essential for star formation studies. 
The molecular cloud core is the densest part of the hierarchical structure of the interstellar medium.
Although there are numerous molecular cloud cores, 
only a small proportion are likely close to the onset of star formation.  
Different initial conditions probably lead to different star formation modes.
Therefore, we need to have good examples in different environments 
[low-mass star formation in less-turbulent ``nearby dark clouds'' vs high-mass star formation in
more turbulent ``giant molecular clouds'' (GMCs)].  
L1544 has served as a very interesting starless core showing gravitational collapse
in nearby dark clouds 
\citep{1998ApJ...504..900T,2019A&A...629A..15R,2019ApJ...874...89C}
that is on the verge of star formation.  
Because the majority of stars form in GMCs rather than dark clouds, 
however, understanding star formation in GMCs is of great importance.
Studies in Infrared Dark Clouds (IRDCs), which 
are dense parts
of GMCs, have been extensively made
\citep{2012ApJ...756...60S,2016A&A...592A..21F,2018ApJ...855....9L,2018ApJ...861...14C,2019ApJ...875...24C,2019ApJ...886..102S},
while the importance of the nearest, archetypal GMC, Orion A, remains because of its proximity.
According to \cite{1990ASPC...12..273B}, 
global properties of GMCs are: masses of $1-2 \times 10^5 \nom{M}$,
mean diameters of 45 pc, and mean average H$_2$ column densities of $3-6 \times 10^{21}$cm$^{-2}$.
IRDCs are clumps inside GMCs with temperatures 10$-$25 K, sizes of 1$-$10 pc, H$_2$ densities of $\sim 10^6$ cm$^{-3}$, and H$_2$ column densities up to  $\sim 10^{23}$ cm$^{-2}$ \citep{1996A&A...315L.165P,2000ApJ...543L.157C}.

It is not clear yet how stars start to form. 
Most observed molecular cloud cores seem to be 
in stable equilibrium 
or have evolutionary timescales much longer than the free-fall timescale,
at least in low-mass star-forming regions \citep{1983ApJ...264..517M,2002ApJ...575..950O}.
Furthermore, molecular cloud cores should be marginally magnetically supercritical in theory \citep{1998ApJ...494..587N}
and seem so observationally \citep{2012ARA&A..50...29C}.
On the other hand, if molecular cloud cores are formed in a non-equilibrium stable state or unstable state 
(even including magnetic contributions), 
they will evolve very quickly,
namely, the timescale to form stars is not much longer than the free-fall time.
Given that observed starless cores are in dynamical equilibrium and stable, 
there should be a mechanism to change a stable core into an unstable one.

N$_2$D$^+$ is one of the most important molecules, as it allows us to investigate
the densest regions of molecular cloud cores
\citep{2017ApJ...834..193K,2017A&A...606A.125S,2018ApJ...855..119A,2019MNRAS.486.4114R,2018A&A...615A..83V,2018A&A...617A..27P,
2018A&A...617A.120M,2019ApJ...870...81T,2019PASJ...71...73T}.
N$_2$H$^+$ is known to be a molecule weakly affected by depletion in cold, dense regions \citep{2002ApJ...570L.101B}.
N$_2$D$^+$ seems to be even less affected by depletion than N$_2$H$^+$ \citep{2007A&A...467..179P}.
Furthermore, the deuterium fraction of some molecular species such as N$_2$D$^+$ increases in the starless phase 
\citep{2005ApJ...619..379C,2006AA...460..709F,2009A&A...496..731E}, 
and then
decreases after the
onset of star formation 
\citep{2012ApJ...747..140S,2018ApJS..237...22S,2018MNRAS.476.1982D}.
Therefore, molecular cloud cores with high deuterium fractions of N$_2$D$^+$ are most likely to be close to the onset of star formation. 
By using the protostellar data and the N$_2$D$^+$ data, we can categorize cores not only ``before'' and ``after"" star formation,
but also evaluate evolutionary stages in them such as ``early starless'', ``late starless'' (closer to the onset of star formation),
``early protostellar'' and ``late protostellar''.
In this paper, we adopted the catalogue of \cite{2018ApJS..236...51Y}, and regard that a core is starless if it is not associated with a protostar.


\section{Observations} \label{sec:obs}

\subsection{Source Selection}

To find the best examples of molecular cloud cores with high deuterium fractions in GMCs, 
we conducted an astrochemical census toward the Orion region.  First, we performed an 
extensive study based on 850 $\mu$m observations with the SCUBA-2 camera onboard the JCMT 
toward the Planck Galactic Cold Clumps (PGCCs; 10-20 K) for the JCMT Legacy program ``SCOPE''
\footnote{https://www.eaobservatory.org/jcmt/science/large-programs/scope/}
 (SCUBA-2 Continuum Observations of Pre-protostellar Evolution, PI = Tie Liu; \citealt{2018ApJS..234...28L}).  
Based on the Planck all-sky survey data \citep{2011AA...536A..23P,2016A&A...594A..28P}, we also carried out a series
of line observations of molecular clouds with ground-based radio telescopes for the TOP-SCOPE Planck Cold Clump collaboration to
understand the initial condition for star formation. 
For example, ``TOP''
\footnote{http://radio.kasi.re.kr/trao/key\_science.php} 
(TRAO Observations of Planck cold clump, PI Tie Liu)
is a related Key Science Program with the TRAO 14-m radio telescope.
To minimize the distance uncertainty, we selected the Orion star-forming region for this study.
For the Orion region, we catalogued 119 SCUBA-2 cores in PGCCs \citep{2018ApJS..236...51Y}.  
Then, we observed these cores with the Nobeyama 45-m telescope in the $J$ = 1$-$0 transitions of 
N$_2$H$^+$, N$_2$D$^+$, DNC, HN$^{13}$C, and other lines \citep{kim2020}. 
From these, we selected two cores bright in N$_2$D$^+$ emission.
One is a star-forming core (G210.82$-$19.47North1; here we call it G210), 
and the other is a starless core  (G211.16-19.33North3; here we call it G211).  
G210 includes the Herschel source HOPS 157 \citep{2016ApJS..224....5F}, which is classified as a Class I Young Stellar Object (YSO).
Its bolometric luminosity and bolometric temperature are $L_{bol}$ = 3.8 \nom{L} and  $T_{bol}$ = 77.6 K, respectively.
This bolometric temperature is close to the boundary value of 70 K between Class 0 and Class I \citep{1995ApJ...445..377C}.
The best-fit envelope mass within 2500 AU (6$\arcsec$) is 0.4 \nom{M}.
However, \cite{2018ApJS..236...51Y} derived the bolometric luminosity and bolometric temperature of
$L_{bol}$ = 1.1 \nom{L} and  $T_{bol}$ = 213 K, respectively.
\cite{2018ApJS..236...51Y} mainly used WISE data points and missed the mid-infrared ranges, which can cause
overestimation of $T_{bol}$.
The spectral index of HOPS 157 is 1.74 \citep{2016ApJS..224....5F} and 1.22 \citep{2018ApJS..236...51Y}, both of which 
suggest that the source is Class I.
Both G210 and G211 belong to the dark cloud L1641 \citep{1962ApJS....7....1L} of the Orion A GMC \citep{1986ApJ...303..375M}, but outside the Orion Nebula region, 
and are close to each other in the sky (22$\arcmin$ = 3 pc apart). 
Given their proximity, they may be cores caught before and after the onset of star formation in {\it similar environments}.  
Furthermore, they are located along the $^{13}$CO $J$ = 1$-$0 filaments observed by \cite{1998AJ....116..336N}.
G210 is located at the northern tip of filament \#24,
and G211 is located at the intersection of two filaments, \#26 and \#28.
By comparing these SCUBA-2 cores with CS cores cataloged in previous observations, we found the following coincidences.
The CS ($J$ = 2$-$1) core FC-24 \citep{1998ApJS..118..517T} corresponds to a combination of
G210.82$-$19.47 North1 (G210) and  North2.
Meanwhile, the CS ($J$ = 1$-$0) core TUKH101 cataloged in \cite{1993ApJ...404..643T} corresponds to G211.
We will discuss the physical parameters of these single-dish CS cores later.

A distance to L1641 in the Orion A giant molecular cloud of 398$\pm$7 pc from
\citet{2019MNRAS.487.2977G} is adopted for this paper.

\subsection{Nobeyama 45-m radio telescope}

Observations with the 45-m radio telescope
of the Nobeyama Radio Observatory\footnote{Nobeyama Radio Observatory
is a branch of the National Astronomical Observatory of Japan,
National Institutes of Natural Sciences.} were performed in 2017 February, 2018 January, and 2018 February
(proposal IDs: CG161004, LP177001).
We observed eight molecular lines, 
CCS $J_N = 7_6-6_5$ and $J_N = 8_7-7_6$, HC$_3$N $J = 9-8$, N$_2$H$^+$ $J = 1-0$,
HN$^{13}$C $J = 1-0$, DNC $J = 1-0$, N$_2$D$^+$ $J = 1-0$, and 85 GHz cyclic C$_3$H$_2$ $J_{KaKc} = 2_{12}-1_{02}$
by using the receivers TZ1 \citep{2013PASP..125..213A,2013PASP..125..252N}, T70, and FOREST \citep{2016SPIE.9914E..1ZM}.
The employed line frequencies are summarized in Table 1. Please refer to
\cite{2017ApJS..228...12T} for references for frequencies.
The employed receiver backend was the SAM45 digital spectrometer \citep{2012PASJ...64...29K}.

The FWHM (full width at half maximum) beam sizes at 86 GHz with TZ1 and T70 were 18$\farcs$2$\pm$0$\farcs$1 and 18$\farcs$9$\pm$0$\farcs$4, respectively.
The main-beam efficiency $\eta_{mb}$ at 86 GHz with TZ1 was 54$\pm$3\%.
The value of $\eta_{mb}$ with T70 was 43$\pm$4\%.
The FWHM beam sizes at 86 GHz with FOREST was 19$\farcs$0$\pm$0$\farcs$5.
The value of $\eta_{mb}$ at 86 GHz with FOREST was 56$\pm$5\%.
The spectral resolution
was 15.26 kHz  (corresponding to 0.05$-$0.06 km s$^{-1}$) for both the TZ1 and T70 observations.
The spectral resolution of the FOREST observations, however,
was set to a twice larger value, 30.52 kHz  (corresponding to 0.10$-$0.12 km s$^{-1}$)
unintentionally.
Single-pointing observations with TZ1 and T70 (partly with FOREST) were performed in the ON-OFF position-switching mode.
Mapping observations with FOREST were made in the on-the-fly mapping mode \citep{2008PASJ...60..445S}.
The map sizes are 5$\arcmin$ square and  12$\arcmin \times 6\arcmin$ for G210 and G211, respectively.

The observed intensity is reported in terms of the corrected
antenna temperature $T_A^*$.
The main-beam
radiation temperature can be derived as $T_{mb}$ = $T_A^*$/$\eta_{mb}$.
Telescope pointing was established by observing relevant 43-GHz SiO maser sources every $\sim$ 60 min,
and was accurate to $\sim$ 5$\arcsec$. 
The observed data were reduced using the software packages ``NewStar'' and ``NoStar'' of the Nobeyama Radio Observatory,
for single-pointing  and on-the-fly observations, respectively.

\setcounter{table}{0}
\begin{table*}[h!]
\renewcommand{\thetable}{\arabic{table}}
\caption{Molecular Lines Observed with the Nobeyama 45-m Radio Telescope}
\center
\begin{tabular}{llc}
\hline 
\hline 
Line & Frequency & rms Noise Level\footnote{At the spectrometer resolution.}\\
      & (GHz) & (K) \\
\hline 
CCS ($J_N$ = 7$_6-6_5)$&81.505208&0.10\\
CCS ($J_N$ = 8$_7-7_6)$&93.870107&0.10\\
HC$_3$N $(J$ = 9$-$8)&81.8814614&0.09\\
N$_2$H$^+$ ($J$ = 1$-$0)&93.1737767&0.12\\
DNC ($J$ = 1$-$0)&76.3057270&0.09\\
HN$^{13}$C ($J$ = 1$-$0)&87.090859&0.09\\
N$_2$D$^+$ ($J$ = 1$-$0)&77.1096100&0.07\\
cyclic C$_3$H$_2$ ($J_{K_aK_c}$ = 2$_{12}-$1$_{01})$&85.338906&0.09\\
\hline
\end{tabular}
\end{table*}

\subsection{ALMA ACA 7-m Array}

We observed the two selected cores using the 7-m Array of the Atacama Compact Array (ACA; a.k.a, the Morita Array)
\citep{2009PASJ...61....1I},
which is a part of the Atacama Large Millimeter/submillimeter Array (ALMA).
Observations were made on 2017 July 13, 2017 August 19, and 2017 August 2
as a  7-m Supplemental Call program of ALMA Cycle 4 (proposal ID: 2016.2.00058.S).
The ALMA Band 6 receivers were used to image the molecular lines listed in Table 2 as well as the 1.2 mm continuum emission.
Rest frequencies were taken from Splatalogue entries \footnote{https://www.cv.nrao.edu/php/splat/} adopted in 
the ALMA Observing Tool\footnote{almascience.org}
except for DNC and HN$^{13}$C, whose frequencies are not listed in this catalog.
Instead, the rest frequencies for DNC and HN$^{13}$C were taken from the JPL Catalog
\citep{1998JQSRT..60..883P}.
Spectral resolution was 122.1 kHz  (corresponding to 0.14$-$0.17 km s$^{-1}$) for the line emission.
The mosaic observations consist of three telescope pointings.
The bandpass and secondary flux calibrator were J0522-3627, and the phase calibrator was J0501-0159 or J0542-0913.
G210 and G211 were observed with the phase calibrator in a repeated cycle,
and should have the same calibration quality.
The number of antennas employed ranged from eight to ten.
The baseline length ranged from 8.9 to 49 m.

The raw data were calibrated by the Joint ALMA Observatory and East-Asian ALMA Regional Center using the pipeline software of CASA
\citep{2007ASPC..376..127M} version 4.7.2.
Imaging was done by the authors manually using CASA 5.4.0.
The continuum emission was collected from three 1.875-GHz-wide continuum spectrum windows centered at
216.0 GHz, 244.0 GHz, and 267.0 GHz, and also from emission-free channels of the line spectrum windows.
We adopted 1.2 mm (242 GHz) as the representative wavelength (frequency) for the continuum emission.
The synthesized beam size of the 1.2 mm continuum image was
$7\farcs26\times3\farcs70$ with a position angle of 88$\fdg$20 for G210,
and
$7\farcs30\times3\farcs70$ with a position angle of 87$\fdg$78 for G211.
The FWHM primary beam size was 42$\arcsec$.
The task ``clean'' of CASA was used for image deconvolution by using Briggs weighting with a `robust' parameter of 0.5.

The intensity is expressed in terms of the brightness temperature $T_b$.
The rms noise level is 2.71 mJy beam$^{-1}$ for continuum,
170 mJy beam$^{-1}$ for CS, H$^{13}$CN, H$^{13}$CO$^+$, and HN$^{13}$C,
and 
130 mJy beam$^{-1}$ for the other molecular lines.
Note that the primary beam response is not corrected for maps,
but is corrected for the spectrum, flux density, line parameter, and physical parameters such as the column density.

\setcounter{table}{1}
\begin{table*}[h!]
\renewcommand{\thetable}{\arabic{table}}
\caption{Molecular Lines Observed with the ACA}
\center
\begin{tabular}{ll}
\hline 
\hline 
Line & Frequency \\
      & (GHz) \\
\hline 
DCO$^+$       ($J$ = 3$-$2)                & 216.112580\\
DCN             ($J$ = 3$-$2)                & 217.238530\\
DNC             ($J$ = 3$-$2)                & 228.910490\footnote{This should be 228.910489, but the difference is negligible for the purposes of this study.}\\
CO               ($J$ = 2$-$1)                & 230.538000\\
N$_2$D$^+$    ($J$ = 3$-$2)                & 231.321828\\
CS               ($J$ = 5$-$4)                 & 244.935556\\
H$^{13}$CN      ($J$ = 3$-$2)                & 259.011787\\
H$^{13}$CO$^+$ ($J$ = 3$-$2)                & 260.255342\\
HCN             ($J$ = 3$-$2)                 & 265.886431\\
HCO$^+$        ($J$ = 3$-$2)                & 267.557633\\
\hline
\end{tabular}
\end{table*}

\section{Results} \label{sec:results}

\subsection{Results from Nobeyama observations}

Table 3 lists the line parameters obtained from observations with the Nobeyama 45-m telescope.  
For N$_2$D$^+$ and N$_2$H$^+$, we show the results of hyperfine fitting.
Both cores have bright emission from the deuterated molecules DNC ($J$ = 1$-$0) and N$_2$D$^+$ ($J$ = 1$-$0)
which characterizes them as being close to the onset of star formation. 
From our earlier sample of 119 cores in Orion, the N$_2$D$^+$ ($J$ = 1$-$0),  detection rate was 49\%,
and typical antenna temperature was $T_A^*$ is 0.1$-$0.3 K.
G210 and G211 are two bright cores in N$_2$D$^+$ emission
with peak antenna temperature of 0.45 K and 0.37 K for G210 and G211, respectively.
For context, the brightest N$_2$D$^+$ core in Orion has  $T_A^*$ (N$_2$D$^+$) = 0.8 K.
Figure 1 shows the N$_2$D$^+$ ($J$ = 1$-$0) line spectra observed toward G210 and G211.
The full-width at half maximum (FWHM) linewidths $\Delta v$ in N$_2$D$^+$ ($J$ = 1$-$0) for G210 and G211 were derived to be 
0.41 km s$^{-1}$ and 0.45 km s$^{-1}$, respectively, by fitting the hyperfine structure. 
These values are much narrower than the width of the deuterated molecular line toward massive protostellar objects (0.7$-$3 km s$^{-1}$) \citep{2006AA...460..709F}.
Note that the linewidth of HN$^{13}$C and DNC is larger than those of other lines, 
because we ignore their hyperfine structure \citep{2007MNRAS.382..840F,2009A&A...507..347V}
when fitting.
Observations of other cores are reported in \citet{kim2020}.

Figure 2 shows the N$_2$H$^+$ ($J$ = 1$-$0) integrated-intensity maps 
toward G210 and G211.
Each map includes a chain of SCUBA-2 cores.
Also, G210.82$-$19.47 North1 (G210), North2, and South are located along 
the $^{13}$CO filament \#24 of \cite{1998AJ....116..336N}.
G211.16-19.33 North1, North2, North 3 (G211), North4, and North5 are located along filament \#26, while
G211.16-19.33 North3 (G211) and South are located along filament \#28.
Therefore, G211.16-19.33 North 3 (G211) is located around the intersection of filaments \#26 and \#28.
Since the velocity difference between these two filaments is only 0.9 km s$^{-1}$, and given the fact that 
G211 is located at this the intersection, the filaments are likely physically connected.
The N$_2$H$^+$ emission observed traces the backbone of these $^{13}$CO filaments.

The CS ($J$ = 2$-$1) core FC-24 \citep{1998ApJS..118..517T}, corresponding to a combination of
G210.82$-$19.47 North1 (G210) and  North2,
has a linewidth of 2.06 km s$^{-1}$ (with a 50$\arcsec$ beam)
and a half width at half maximum (HWHM) radius of 0.40 pc in CS ($J$ = 2$-$1).
The CS ($J$ = 1$-$0) core TUKH101 \citep{1993ApJ...404..643T}, corresponding to G211,
has a linewidth of 0.99 km s$^{-1}$ (with a 36$\arcsec$ beam)
and an HWHM radius of 0.15 pc in CS ($J$ = 1$-$0).
The linewidth in N$_2$D$^+$ toward G211 has been measured as
0.37 km s$^{-1}$,  and is much narrower than the linewidth in CS
by a factor of 2.7.
It is likely that the non-thermal turbulent motions observed in the larger, less-dense CS core, 
have been largely dissipated in the high-density core center traced by N$_2$D$^+$ emission.

\setcounter{table}{2}
\begin{table*}[h!]
\renewcommand{\thetable}{\arabic{table}}
\caption{Line parameters from Observations with the Nobeyama 45-m Telescope}
\center
\begin{tabular}{lllllll}
\hline 
\hline Source &	Line								&	$T_A^*$		&	$V_{LSR}$	&	$\Delta v$\footnote{DNC and HN$^{13}$C linewidths are overestimated, because we ignore hyperfine splitting.}	&	$T_{ex} * \tau$(main)\footnote{The optical depth of the main (brightest) hyperfine component.}			&	$\tau$(main)	\\
		&									&	K			&	km s$^{-1}$	&	km s$^{-1}$	&	K						&				\\
\hline 
G210	&	82 GHz CCS ($J_N$ = 7$_6-6_5$)				&		$<$	0.18	&	. . .		&	. . .		&	. . .						&	. . .			\\
	&	94 GHz CCS ($J_N$ = 8$_7-6_5$)				&		$<$	0.19	&	. . .		&	. . .		&	. . .						&	. . . 			\\
	&	HC$_3$N ($J$ = 9$-$8)						&		$<$	0.22	&	. . .		&	. . .		&	. . .						&	. . .			\\
	&	c-C$_3$H$_2$ ($J_{K_aK_c}$ = 2$_{12}-$1$_{01}$)	&	0.47 			&	5.31 		&	0.76 		&	. . .						&	. . .			\\
	&	DNC ($J$ = 1$-$0)						&	0.82 			&	5.43 		&	0.98		&	. . .			&	. . .			\\
	&	HN$^{13}$C	($J$ = 1$-$0)					&	0.21 			&	5.47 		&	1.12		&	. . .						&	. . .			\\
	&	N$_2$D$^+$ ($J$ = 1$-$0)					&	0.45 			&	5.34 		&	0.41		&	3.4 	$\pm$	0.2 			&	0.4$\pm$0.1 			\\
	&	N$_2$H$^+$ ($J$ = 1$-$0)					&	1.30 			&	5.30 		&	0.44		&	13.5 	$\pm$	0.9 			&	2.7$\pm$0.7 			\\
G211	&	82 GHz CCS ($J_N$ = 7$_6-6_5$)				&	0.18 			&	3.28 		&	0.40		&	. . .						&	. . .			\\
	&	94 GHz CCS ($J_N$ = 8$_7-6_5$)				&		$<$	0.15	&	. . .		&	. . .		&	. . .						&	. . .			\\
	&	HC$_3$N ($J$ = 9$-$8)						&	0.70 			&	3.39 		&	0.39		&	. . .						&	. . .			\\
	&	c-C$_3$H$_2$ ($J_{K_aK_c}$ = 2$_{12}-$1$_{01}$)	&	1.34 			&	3.25 		&	0.43 		&	. . .						&	. . .			\\
	&	DNC ($J$ = 1$-$0)						&	0.96 			&	3.48 		&	1.09		&	. . .       		&	. . .			\\
	&	HN$^{13}$C ($J$ = 1$-$0)					&	0.46 			&	3.36 		&	0.68		&	. . .						&	. . .			\\
	&	N$_2$D$^+$ ($J$ = 1$-$0)					&	0.37 			&	3.38 		&	0.45		&	3.0 	$\pm$	0.6 			&	1.1$\pm$1.1 			\\
	&	N$_2$H$^+$ ($J$ = 1$-$0)					&	1.03 			&	3.33 		&	0.35		&	19.5 	$\pm$	1.4 			&	9.9$\pm$1.2 			\\

\hline
\end{tabular}
\tablecomments{The error of the line intensity is about 10\%. The errors of the LSR velocity and linewidth are less than 0.1km s$^{-1}$.}
\end{table*}

\begin{figure*}
\includegraphics[bb=0 0 800 800, width=10cm]{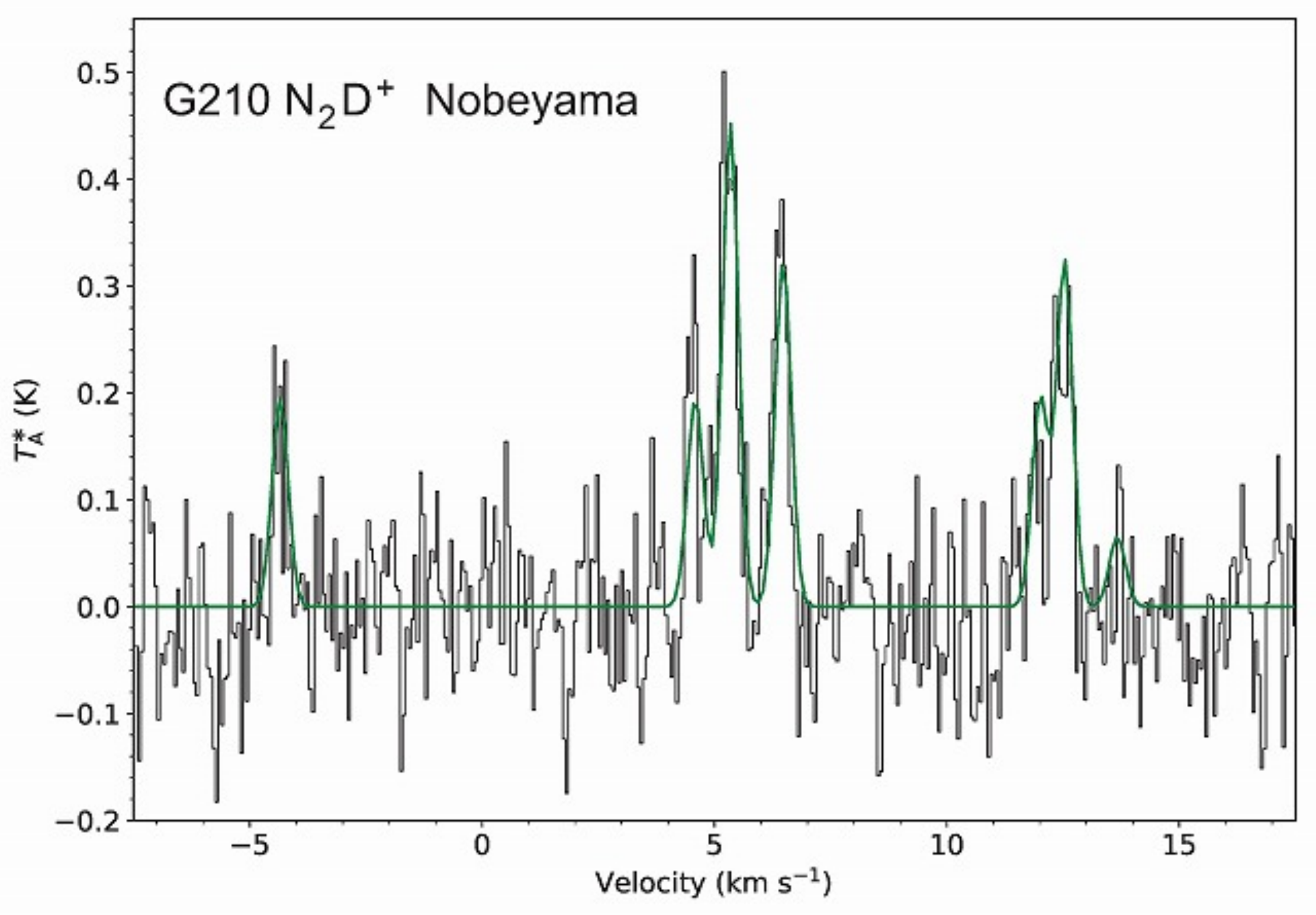}
\includegraphics[bb=0 0 800 800, width=10cm]{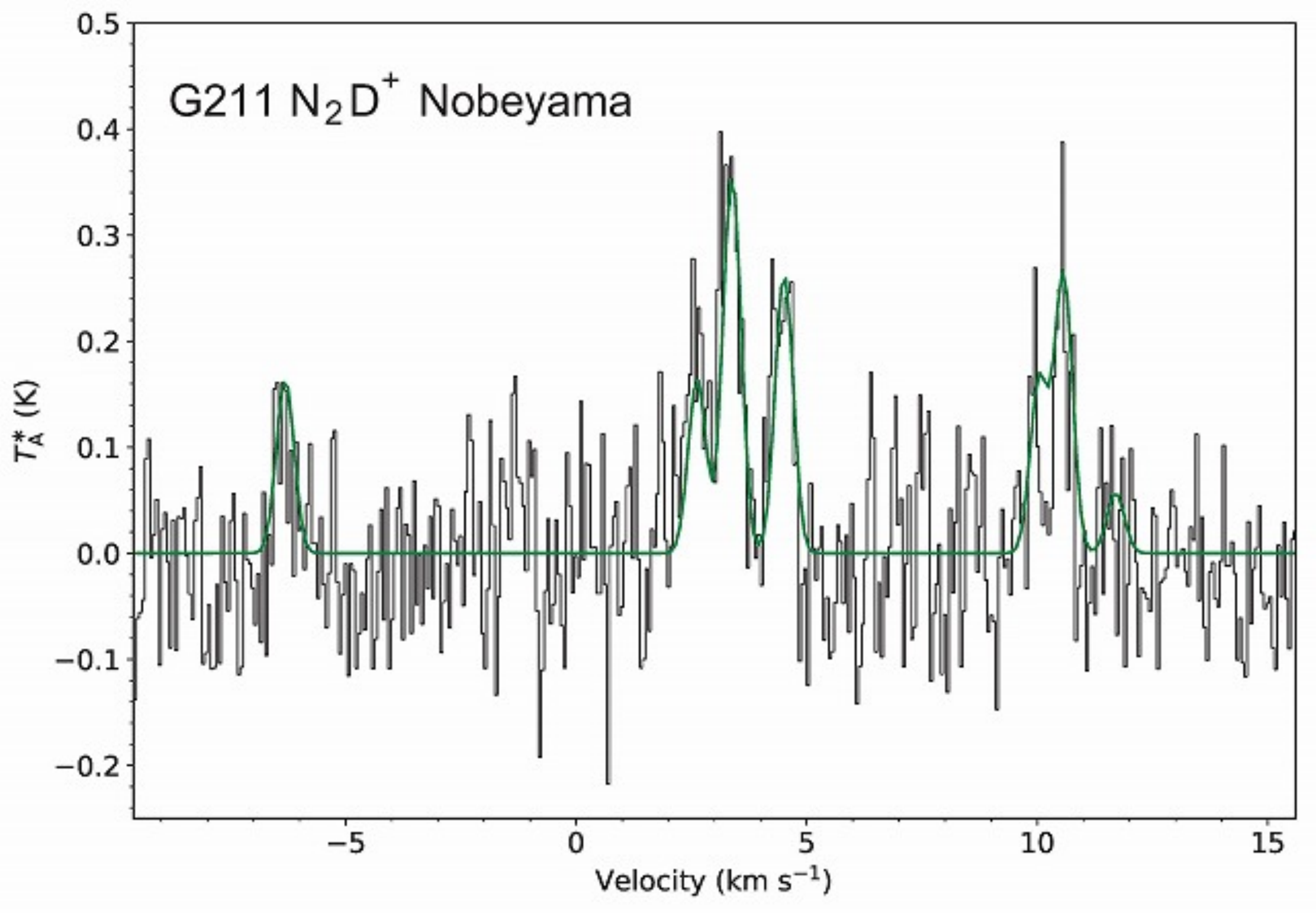}
\caption{N$_2$D$^+$ spectra (black lines) obtained toward G210 and G211 with the Nobeyama 45-m radio telescope and hyperfine fitting results (smooth green curves).
(left) G210 and (right) G211.
}
\end{figure*}

\begin{figure*}
\includegraphics[bb=0 0 800 800, width=10cm]{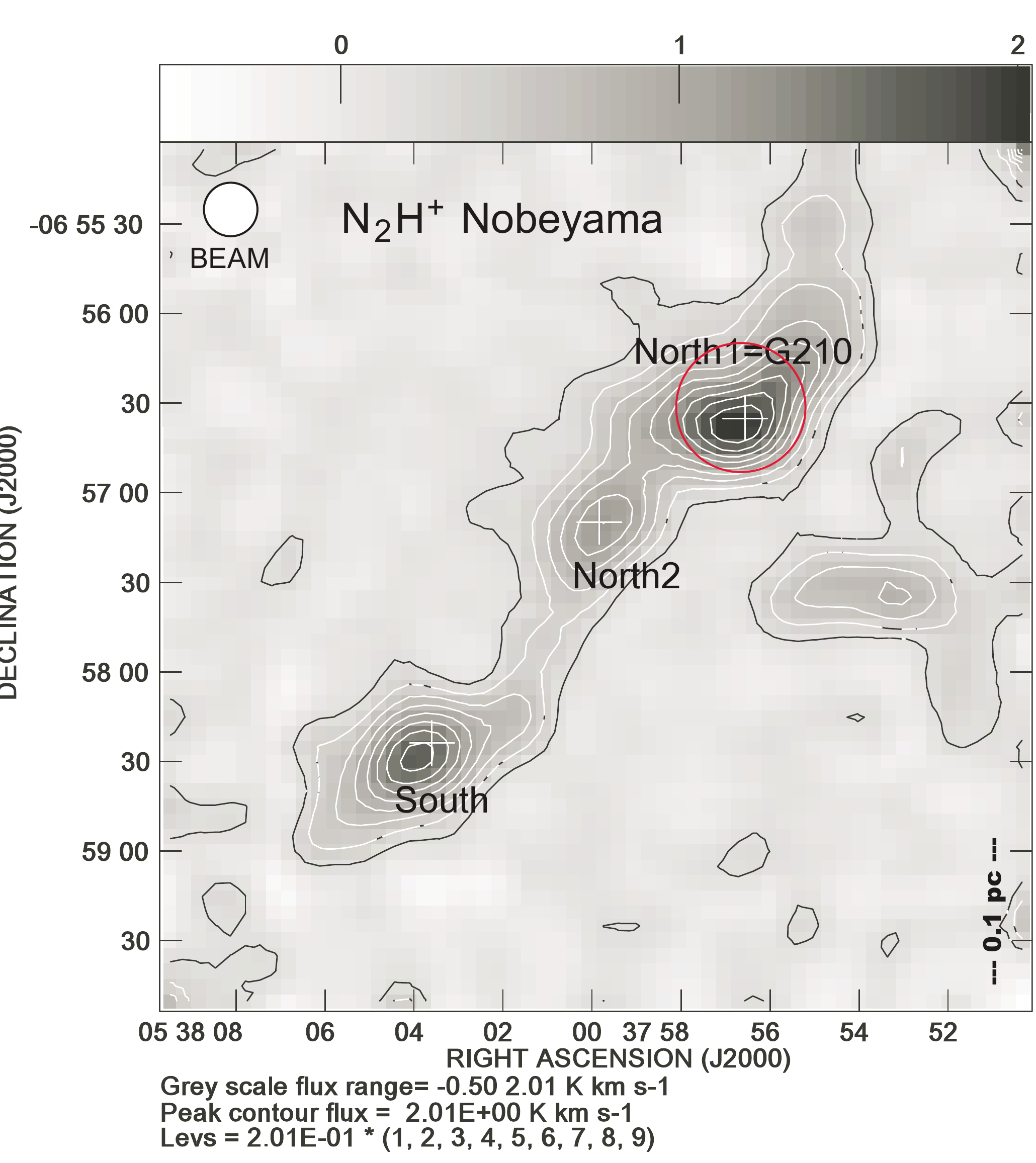}
\includegraphics[bb=0 0 800 800, width=10cm]{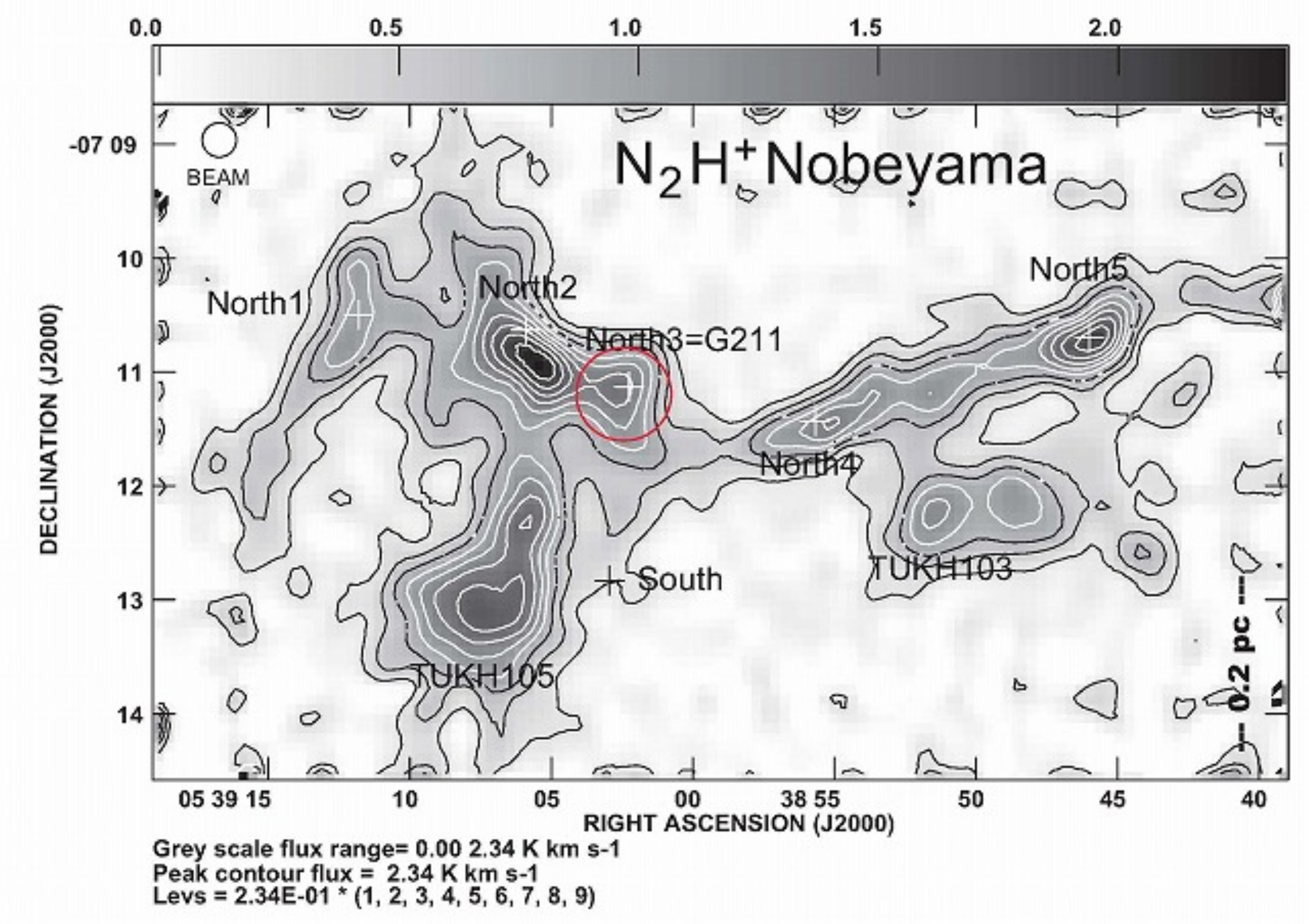}

\caption{The integrated intensity map of N$_2$H$^+$ $J$ = 1$-$0 toward the G210.82$-$19.47 region including G210
and  the G211.16-19.33 region including G211 obtained with the Nobeyama 45-m telescope.
The emission is integrated for the main hyperfine component group $J = 1\rightarrow0$ 
$F_1$ = 2$\rightarrow$1 consisting of three components including the most intense one.
Plus signs represent the locations of SCUBA-2 peak positions.
The red circle represents the field of view (50\% level) of the corresponding ACA observations.
}
\end{figure*}

The column densities are calculated using the method described in \cite{2017ApJS..228...12T}.
Excitation temperatures are derived for N$_2$D$^+$ and N$_2$H$^+$ through hyperfine component fitting,
and that for the other molecules was assumed to be 8 K,
which is half the typical dust temperature for PGCCs (16 K).
The ratio of the excitation temperature to the gas kinetic temperature (assumed to be equal to the dust temperature) 
for the $J$ = 1$-$0 molecular lines
was assumed to be 50\% from Figure 13 of \cite{2008PASJ...60..407T}.
The deuterium fraction $N$(N$_2$D$^+$)/$N$(N$_2$H$^+$)
was derived as 
0.32$\pm$0.03 and 0.24$\pm$0.05 for G210 and G211, respectively.
For Orion cores, $N$(N$_2$D$^+$)/$N$(N$_2$H$^+$) ranges from 0.04 to 0.4 with a median of  0.2 for star-forming cores,
and  from 0.06 to 0.4 with a median of 0.2 for starless cores.
$N$(DNC)/$N$(HN$^{13}$C) is  4.0$\pm$2.8 and 3.9$\pm$0.8 for G210 and G211, respectively,
which correspond to D/H = 0.067$\pm$0.047 and 0.065$\pm$0.01 by assuming $^{12}$C/$^{13}$C = 60 \citep{2002ApJ...578..211S}.
For Orion cores, $N$(DNC)/$N$(HN$^{13}$C) ranges from 1.0 to 13.7 with a median of 2.7 for star-forming cores,
and from 0.7 to 7.7 with a median of 3.8 for starless cores \citep{kim2020}.
G210 and G211 are relatively high deuterium fraction cores in Orion.

\bigskip
\bigskip

\subsection{Results from the ACA observations}

Figure 3 shows that in the 1.2 mm continuum
G210 exhibits only a single peak while
G211 exhibits five emission peaks in the fields of view.  
The G210 dust  continuum core is associated with the {\it Herschel} protostar HOPS 157.  
The primary-beam-corrected 1.2 mm continuum flux density is 69 mJy,
measured by fitting a two-dimensional Gaussian to the map using the task ``imfit'' in CASA.
The SCUBA-2 flux density at 850 $\mu$m is 713 mJy with a beam size of 14$\arcsec$.
If we assume a spectral index of 
$\beta$ = 1.8 \citep{2014A&A...566A..45L},
41\% of the dust emission is recovered by the ACA observations. 
For G211, the sum of the primary-beam-corrected 1.2 mm continuum flux densities of peaks G211A$-$E is 14 mJy.
The SCUBA-2 flux density at 850 $\mu$m is 281 mJy with a beam size of 14$\arcsec$.
Again, if we assume a spectral index of 
$\beta$ = 1.8 \citep{2014A&A...566A..45L},
21\% of the dust emission is recovered by the ACA observations.

In addition to the 1.2 mm continuum, G210 was detected in 
N$_2$D$^+$ ($J$ = 3$-$2), DNC ($J$ = 3$-$2), DCN ($J$ = 3$-$2), DCO$^+$ ($J$ = 3$-$2), 
HCO$^+$ ($J$ = 3$-$2), CS ($J$ = 5$-$4), HCN ($J$ = 3$-$2), HCO$^+$ ($J$ = 3$-$2), 
H$^{13}$CO$^+$ ($J$ = 3$-$2), H$^{13}$CN ($J$ = 3$-$2), and CO ($J$ = 2$-$1) 
with the ACA 7m Array.
Integrated intensity maps are shown in Figures 4 and 5 (the CO distribution will be shown later). 
Note that the N$_2$D$^+$ map shows two symmetric peaks centered 
on the dust continuum peak (plus).  
The two peaks have similar intensities with the weaker peak being 60-70\% as bright as the stronger one.
The CS ($J$ = 5$-$4), H$^{13}$CO$^+$ ($J$ = 3$-$2), and
H$^{13}$CN ($J$ = 3$-$2) integrated intensities, however, are concentrated toward the dust continuum peak.
The DCO$^+$ ($J$ = 3$-$2) and DNC ($J$ = 3$-$2) integrated intensities show a hint of the double peaks, but
the brightness of the two peaks are rather different.
Furthermore, the peak positions in these latter lines are different from those of the N$_2$D$^+$ peaks.
Molecules that are not deuterated do not show the double peaks,
and likely represent the underlying column density and temperature distribution,
which increase toward the position of the protostar, in a simple manner.

\begin{figure*}
\includegraphics[bb=0 0 800 800, width=10cm]{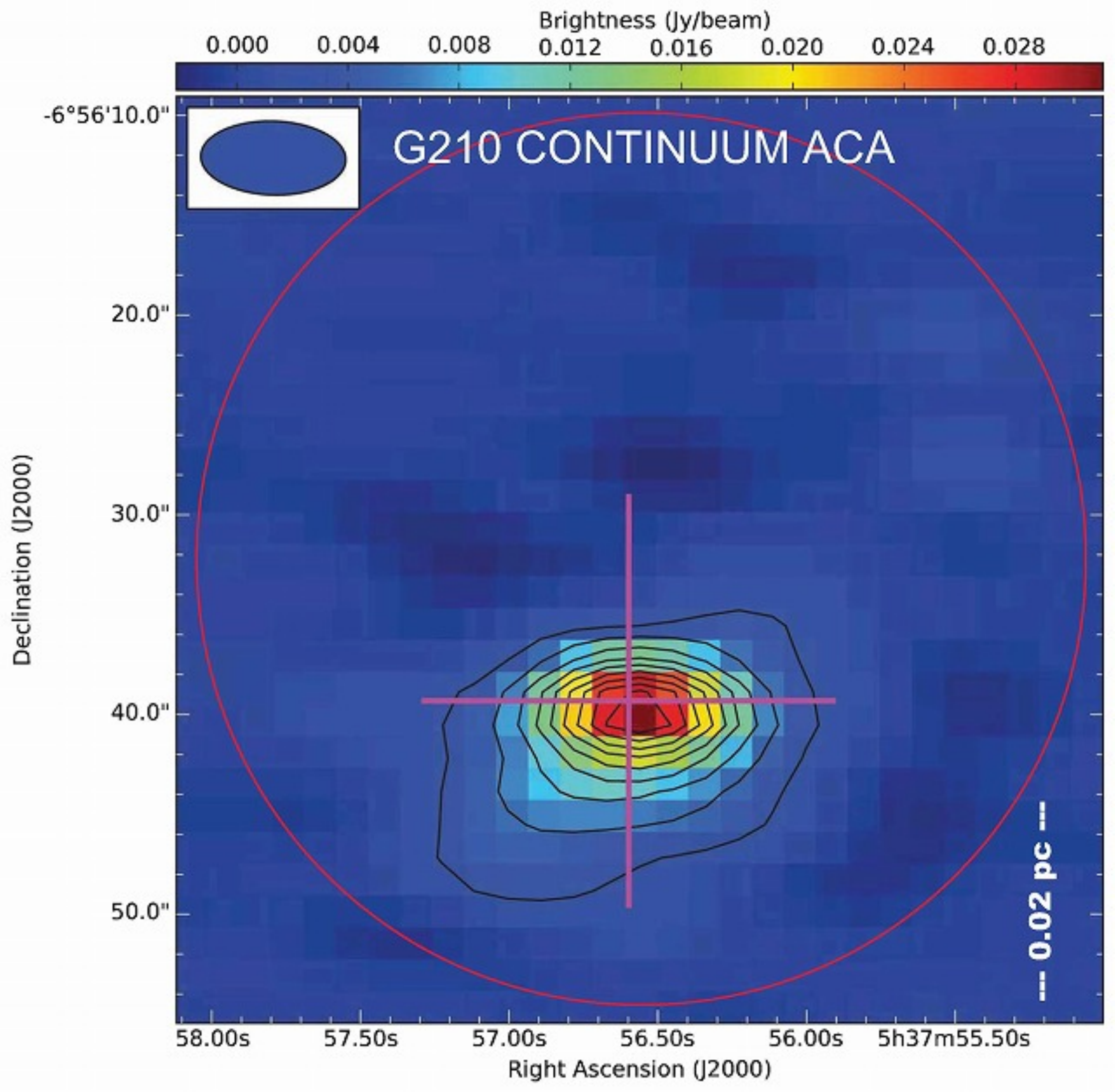}
\includegraphics[bb=0 0 800 800, width=10cm]{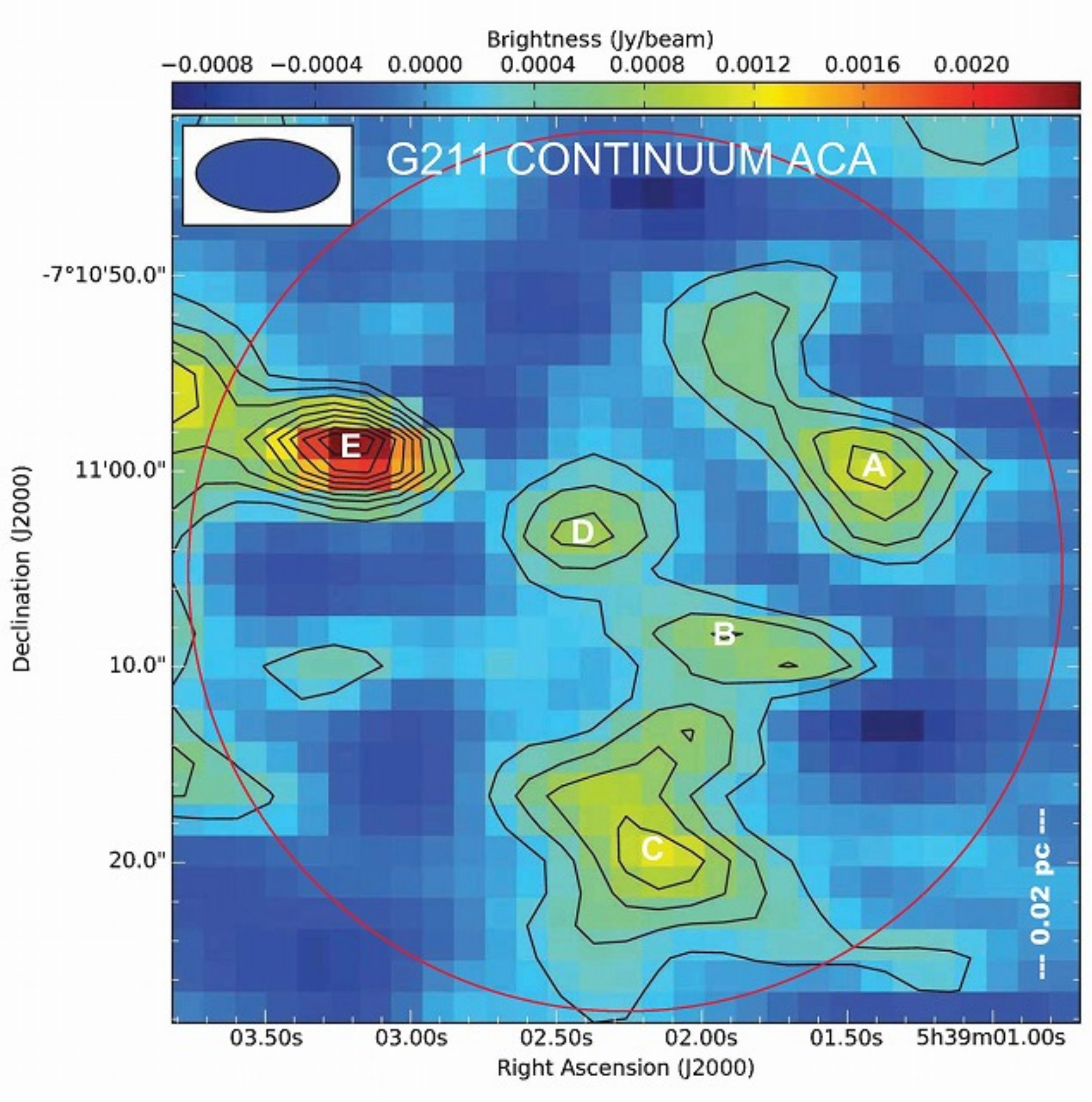}
\caption{
The 1.2 mm continuum map toward G210 and G211.
Contour levels are
shown in steps of 10\% of the maximum value.
The cross symbol represents the continuum peak position of G210.
Local peaks in G211 are designated A through E.
The ellipse in the top-left corner represents the half-intensity synthesized beam size.
The red circle illustrates the field of view (50\% level) of the ACA observations.
}
\end{figure*}

\begin{figure*}
\includegraphics[bb=0 0 600 600, width=20cm]{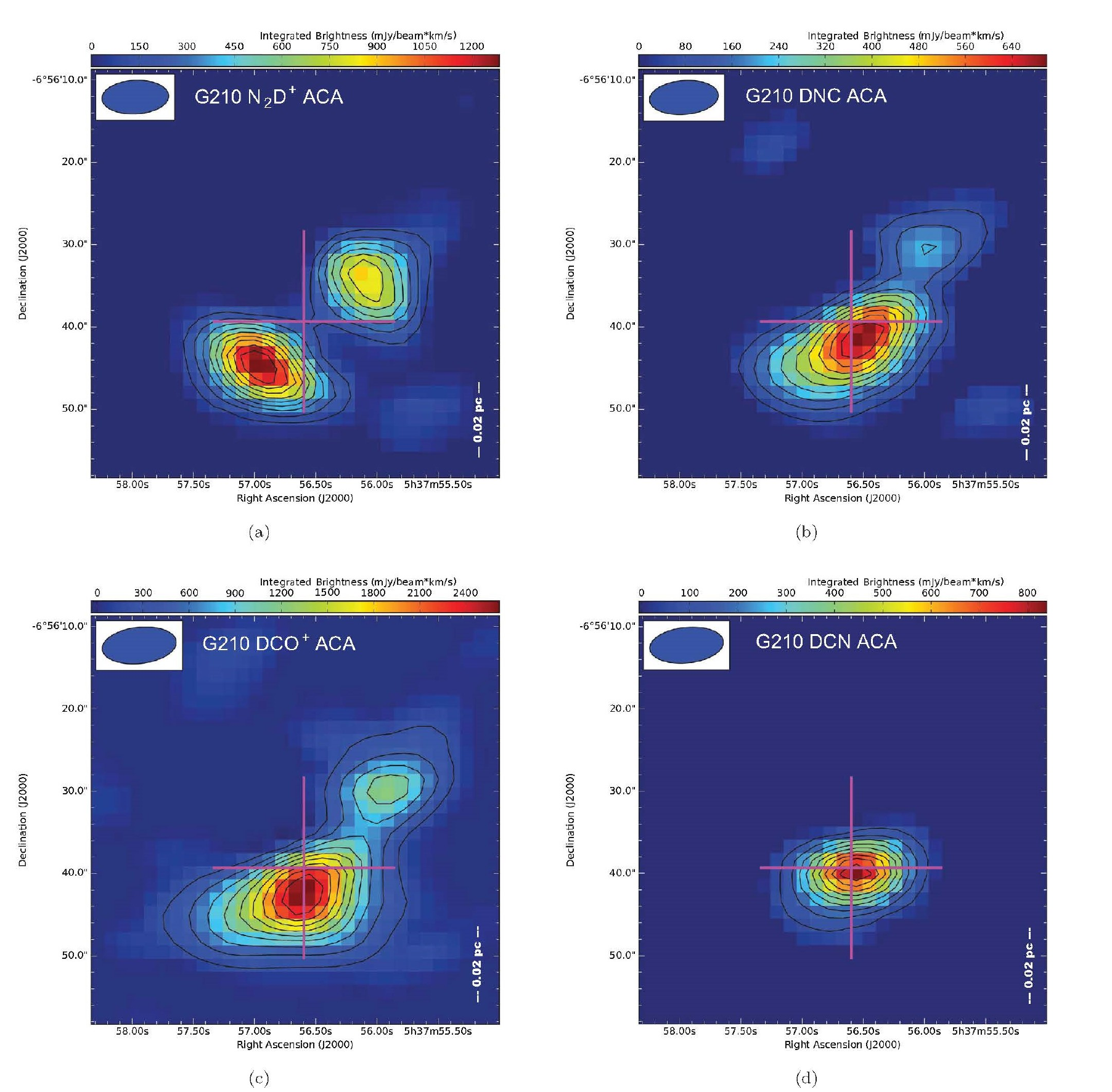}
\caption{The velocity-integrated intensity maps of G210 in
(a) N$_2$D$^+$ $J$ = 3$-$2 ($V_{LSR}$ = 4.57$-$5.94 km s$^{-1}$),
(b) DNC $J$ = 3$-$2 ($V_{LSR}$ = 4.81$-$5.75 km s$^{-1}$),
(c) DCO$^+$ $J$ = 3$-$2 ($V_{LSR}$ = 4.59 to 6.00 km s$^{-1}$),
and
(d) DCN $J$ = 3$-$2 ($V_{LSR}$ = 4.81$-$6.11 km s$^{-1}$).
The N$_2$D$^+$ emission is integrated for the main hyperfine component group
including the brightest one.
1 mJy beam$^{-1}$ corresponds to 0.69 mK, 0.67 mK, 0.66 mK, and 0.67 mK for (a), (b), (c), and (d), respectively.
}
\end{figure*}

\begin{figure*}
\includegraphics[bb=0 0 600 600, width=20cm]{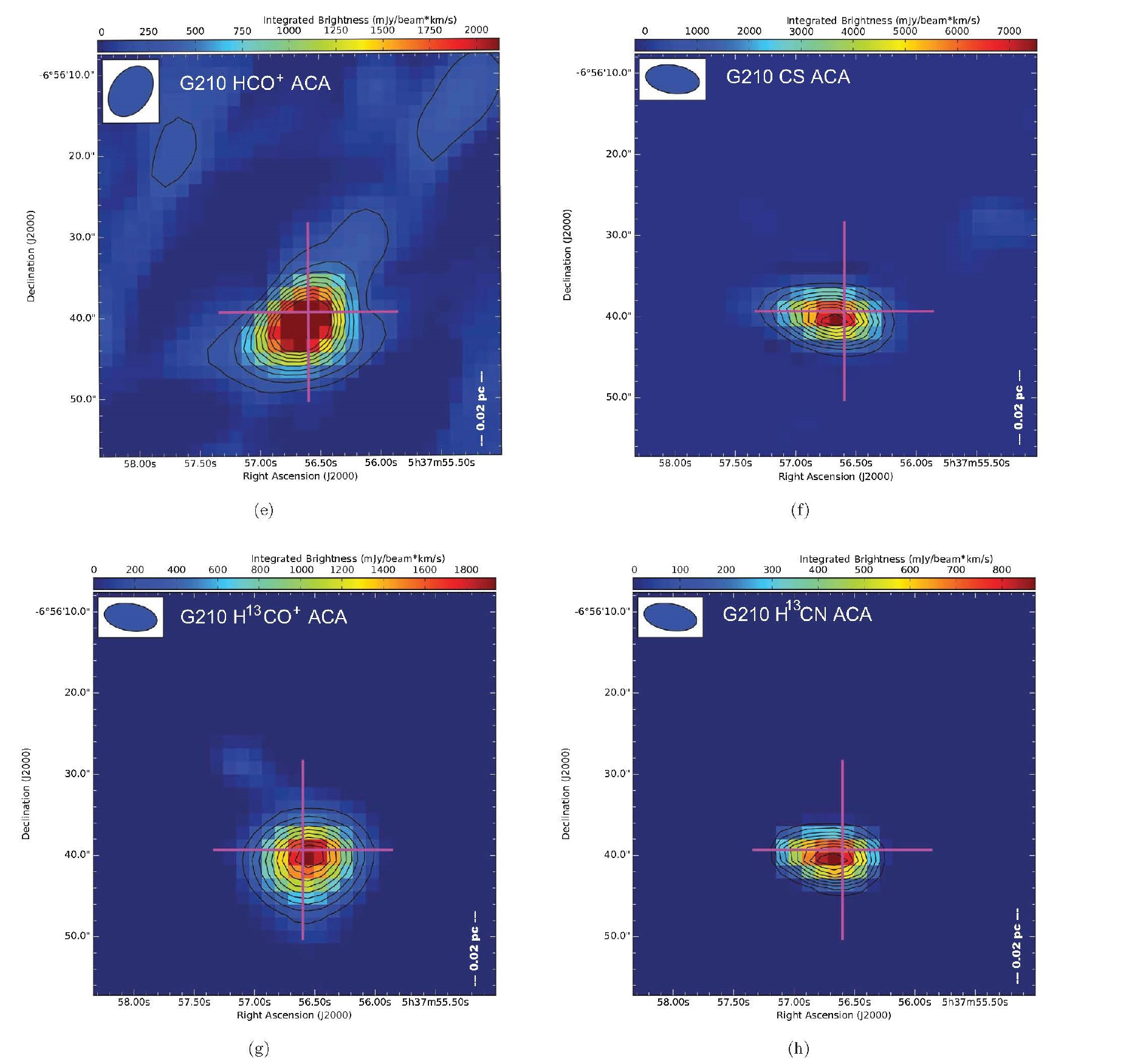}
\caption{Information is similar Figure 4 showing the velocity-integrated intensity maps of G210 in
(e) HCO$^+$ $J$ = 3$-$2 ($V_{LSR}$ = 4.94$-$5.41 km s$^{-1}$),
(f) CS $J$ = 5$-$4 ($V_{LSR}$ = 3.71$-$.53 km s$^{-1}$),
(g) H$^{13}$CO$^+$ $J$ = 3$-$2 ($V_{LSR}$ = 4.60$-$6.03 km s$^{-1}$), and
(h) H$^{13}$CN $J$ = 3$-$2 ($V_{LSR}$ = 4.67$-$6.72 km s$^{-1}$).
1 mJy beam$^{-1}$ corresponds to 0.52, 0.88, 0.83, and 0.85 mK for (e), (f), (g), and (h), respectively.
}
\end{figure*}

G211 was detected in DCO$^+$ ($J$ = 3$-$2), HCO$^+$ ($J$ = 3$-$2),
DNC ($J$ = 3$-$2), N$_2$D$^+$ ($J$ = 3$-$2), and CO ($J$ = 2$-$1), but
was not detected in the other lines.

Figures 6 and 7 show the integrated intensity maps of G211 in various
molecular lines.
The distribution of these molecular lines is very different from the dust continuum distribution.
DCO$^+$ and
HCO$^+$
have intensity maxima toward G211A.
N$_2$D$^+$ and
DNC 
have intensity maxima near G211C, but
are displaced toward G211B.
G211E is not bright but is weakly detected in 
HCO$^+$, 
N$_2$D$^+$, and
DNC.
We see appreciable differences in molecular line distribution, which most likely represents distinct chemical variation.
The distribution of DCO$^+$ and
HCO$^+$
are similar, and
that of 
N$_2$D$^+$ and
DNC 
are also similar.
The CO distribution (Fig. 7) has local peaks
corresponding to G211A and C.

\begin{figure*}
\includegraphics[bb=0 0 600 600, width=20cm]{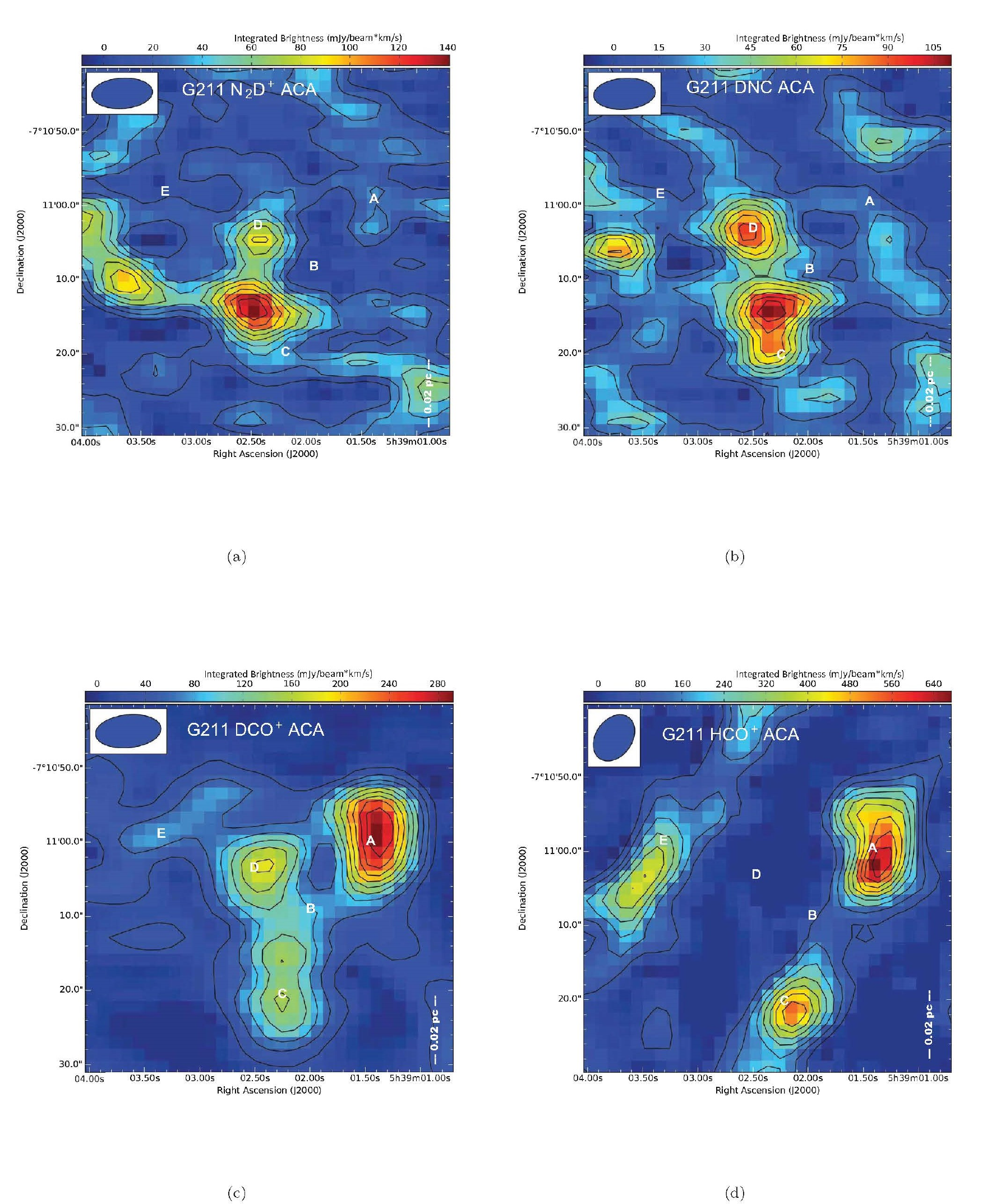}
\caption{Information is similar to Figure 4 and include the velocity-integrated intensity maps of G211 in
(a) N$_2$D$^+$ $J$ = 3$-$2 ($V_{LSR}$ = 2.86$-$4.18 km s$^{-1}$), 
(b) DNC  $J$ = 3$-$2 ($V_{LSR}$ = 3.03$-$4.02 km s$^{-1}$),
(c) DCO$^+$ $J$ = 3$-$2 ($V_{LSR}$ = 3.13$-$3.81 km s$^{-1}$),
and
(d) HCO$^+$ $J$ = 3$-$2 ($V_{LSR}$ = 3.36$-$3.83 km s$^{-1}$).
1 mJy beam$^{-1}$ corresponds to 0.68, 0.67, 0.67, and 0.52 mK for (a), (b), (c), and (d), respectively.
The continuum peak positions are labeled.
}
\end{figure*}

\begin{figure*}
\includegraphics[bb=0 0 800 800, width=20cm]{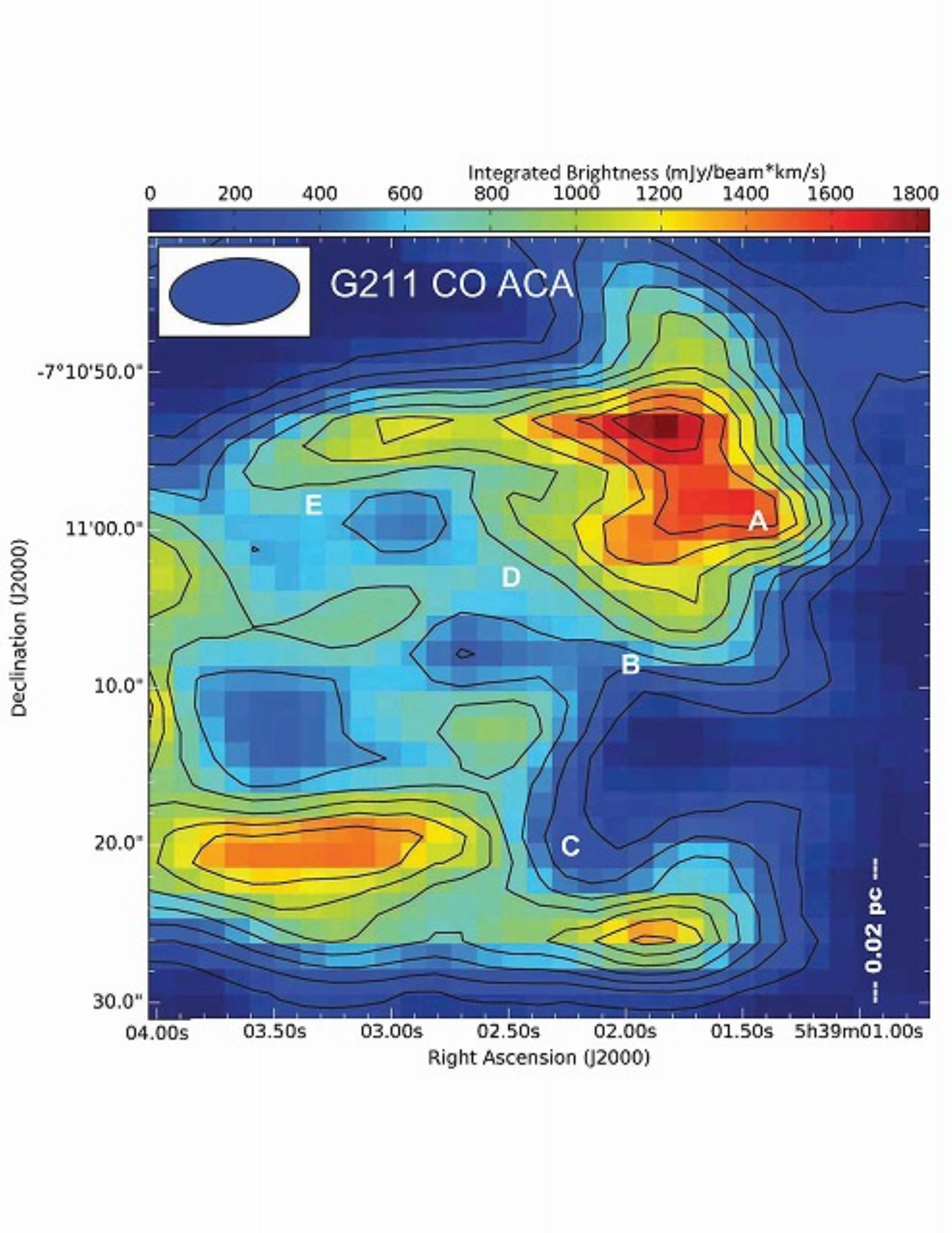}
\caption{
Information is similar to Figure 4 with the velocity-integrated intensity maps of G211 in
CO $J$ = 2$-$1.
1 mJy beam$^{-1}$ corresponds to 0.70 mK.
}
\end{figure*}

\subsection{Observed Parameters from the ACA}

In Tables 4 and 5, we summarize the result of the continuum observations.
Note that the FWHM major and minor axes in Table 4 are not corrected for the synthesized beam size,
while the radius $R$, brightness, and flux density are corrected for the synthesized beam size.
The synthesized beam size is 7$\farcs3\times3\farcs7$.
The catalogued position of HOPS 157 (R.A. = $5^h37^m56\fs57$;  DEC = $-6\degree56\arcmin 39\farcs1$) is almost identical to the G210 continuum source position within 2$\arcsec$.
We calculate the mass $M_{dust}$ from the flux density of the dust continuum emission.
The gas-to-dust mass ratio is assumed to be 100, 
and the opacity of dust grains with thin ice mantles at
a gas density of 10$^6$ cm$^{-3}$ from
\cite{1994A&A...291..943O} is adopted.
The derived dust continuum mass of G210 (0.80 \nom{M}) is twice the {\it Herschel} best-fit envelope mass of the central compact component within 2500 AU or 6$\arcsec$ (0.38 \nom{M})\citep{2016ApJS..224....5F}.
The difference in the flux density ($6.9\times10^{-2}$ Jy) and peak brightness ($2.8\times10^{-2}$ Jy beam$^{-1}$) with a $\sim$ 5$\arcsec$ beam
is also approximately two, which is consistent with the above difference.
The factor of two difference means that there are two components: half of the mass is located in the point-like source, while the other half is located in the envelope.
The masses of the sub-cores in G211 are less than 0.1 $\nom{M}$, and it seems natural that G211 is starless.
These sub-cores need to accrete more mass from outside to form protostars. This topic will be discussed in \S4.3.

\setcounter{table}{3}
\begin{table*}[h!]
\renewcommand{\thetable}{\arabic{table}}
\caption{Continuum Source Coordinate and Size from ACA}
\center
\begin{tabular}{llllrl}
\hline 
\hline 
Core		&	IR association	&	RA (J2000)	&	DEC (J2000)	&	FWHM	&	Source Size	\\
		&		&		&		&	major	&	minor 	\\
		&		&	$^{h\quad m\quad s}$	&	$ \degree\quad \arcmin\quad \arcsec$	&	\quad $\arcsec$	&	\quad $\arcsec$	\\
\hline												
G210		&	star forming	&	5 37 56.6	&	-6 56 40	&	10.3	&	6.5	\\
												
G211	A	&	starless	&	5 39 01.4	&	-7 11 00	&	8.0	&	4.5	\\
G211	B	&	starless	&	5 39 01.7	&	-7 11 10	&	9.5	&	4.1	\\
G211	C	&	starless	&	5 39 02.2	&	-7 11 19	&	11.4	&	9.8	\\
G211	D	&	starless	&	5 39 02.4	&	-7 11 04	&	6.2	&	4.6	\\
G211	E	&	starless	&	5 39 03.2	&	-7 10 59	&	7.9	&	3.6	\\
\hline
\end{tabular}
\tablecomments{The continuum peak position error is estimated to be within 2$\arcsec$.  The source size error is about 3\% for G210 and 10\% for G211.}
\end{table*}

\setcounter{table}{4}
\begin{table*}[h!]
\renewcommand{\thetable}{\arabic{table}}
\caption{Continuum Parameters from ACA}
\center
\begin{tabular}{lllllllll}
\hline 
\hline 
		&		&		&		&		&		\\
Core		&	Brightness	&	Flux Density	&	$R_{dust}$	&	$N$(H$_2$)$_{dust}$	&	$M_{dust}$	\\
		&	Jy beam$^{-1}$	&	Jy	&	pc	&	cm$^{-2}$	&	$\nom{M}$	\\
\hline												
G210		&	2.8E-02	&	6.9E-02	&	0.006 	&	2.8E+23	&	0.72 	\\
G211	(all)	&	8.0E-03	&	1.4E-02	&		&	. . .	&	0.15 	\\
G211	A	&	1.4E-03	&	1.9E-03	&	0.0029 	&	3.3E+22	&	0.019 	\\
G211	B	&	8.2E-04	&	1.2E-03	&	0.0034 	&	1.6E+22	&	0.012 	\\
G211	C	&	1.6E-03	&	6.6E-03	&	0.009 	&	1.3E+22	&	0.07 	\\
G211	D	&	8.3E-04	&	8.8E-04	&	0.0012 	&	8.5E+22	&	0.009 	\\
G211	E	&	3.3E-03	&	3.5E-03	&	0.0012 	&	3.6E+23	&	0.036 	\\
\hline
\end{tabular}
\tablecomments{The flux error is about 10\%. The radius error is about 10\%. The relative errors of the column density and mass are about 30\%.}
\end{table*}

Next, we summarize the line observations with the ACA.
Line intensities at the integrated intensity peak positions and averaged over the areas are listed in Tables 6 and 7, respectively.
Figure 8 depicts the regions where the emission is averaged for spectra.  The map is not primary-beam corrected; however, the brightness
(and the spectra shown later)
are primary-beam corrected to derive physical parameters.

\setcounter{table}{5}
\begin{table*}[h!]
\renewcommand{\thetable}{\arabic{table}}
\caption{ACA Line Parameters at the Intensity Peak Position}
\center
\begin{tabular}{lllll}
\hline 
\hline 
		&	N$_2$D$^+$	&	DNC	&	DCO$^+$	&	HCO$^+$	\\
		&	$T_R$	&	$T_R$	&	$T_R$	&	$T_R$	\\
Core		&	K	&	K	&	K	&	K	\\
\hline										
G210		&	1.80 	&	1.07 	&	3.60 	&	5.20 	\\
										
G211	A	&	. . .	&	. . .	&	0.95 	&	1.90 	\\
G211	C	&	0.17 	&	0.27 	&	0.51 	&	1.50 	\\
G211	D	&	0.22 	&	0.22 	&	0.42 	&	. . .	\\
G211	E	&	0.30 	&	0.22 	&	. . .	&	1.15 	\\
\hline
\end{tabular}
\tablecomments{The intensity error is about 10\%.}
\end{table*}

\setcounter{table}{6}
\begin{table*}[h!]
\renewcommand{\thetable}{\arabic{table}}
\caption{ACA Line Parameters Obtained After Averaging over Area (Figure 8)}
\center
\begin{tabular}{llllllllllllll}
\hline 
\hline 
		&	N$_2$D$^+$	&		&		&		&	DNC	&		&		&	DCO$^+$	&		&		&	HCO$^+$	&		&		\\
		&	$T_R$	&	$\int T_A^* dv$	&	$V_{LSR}$	&	$\Delta v$	&	$T_R$	&	$V_{LSR}$	&	$\Delta v$	&	$T_R$	&	$V_{LSR}$	&	$\Delta v$	&	$T_R$	&	$V_{LSR}$	&	$\Delta v$	\\
Core		&	K	&	K km s$^{-1}$	&	km s$^{-1}$	&	km s$^{-1}$	&	K	&	km s$^{-1}$	&	km s$^{-1}$	&	K	&	km s$^{-1}$	&	km s$^{-1}$	&	K	&	km s$^{-1}$	&	km s$^{-1}$	\\
\hline																												
G210		&	0.28 	&	0.208 	&	5.16 	&	0.69 	&	0.21 	&	5.14 	&	0.49 	&	0.86 	&	5.18 	&	0.55 	&	0.50 	&	5.53 	&	1.19 	\\
																												
G211	A	&	. . .	&	. . .	&	. . .	&	. . .	&	. . .	&	. . .	&	. . .	&	0.74 	&	3.48 	&	0.22 	&	1.03 	&	3.59 	&	0.30 	\\
																												
G211	C	&	0.06 	&	0.023 	&	3.38 	&	0.34 	&	0.11 	&	3.53 	&	0.30 	&	0.25 	&	3.64 	&	0.28 	&	0.23 	&	3.05 	&	0.73 	\\
G211	D	&	0.09 	&	0.030 	&	3.12 	&	0.32 	&	0.11 	&	3.23 	&	0.28 	&	0.31 	&	3.19 	&	0.29 	&	. . .	&	. . .	&	. . .	\\
\hline
\end{tabular}
\tablecomments{The errors of the intensity and linewidth are about 10\%. The LSR velocity error is about 0.1 km s$^{-1}$.}
\end{table*}

\begin{figure*}
\includegraphics[bb=0 0 600 600, width=20cm]{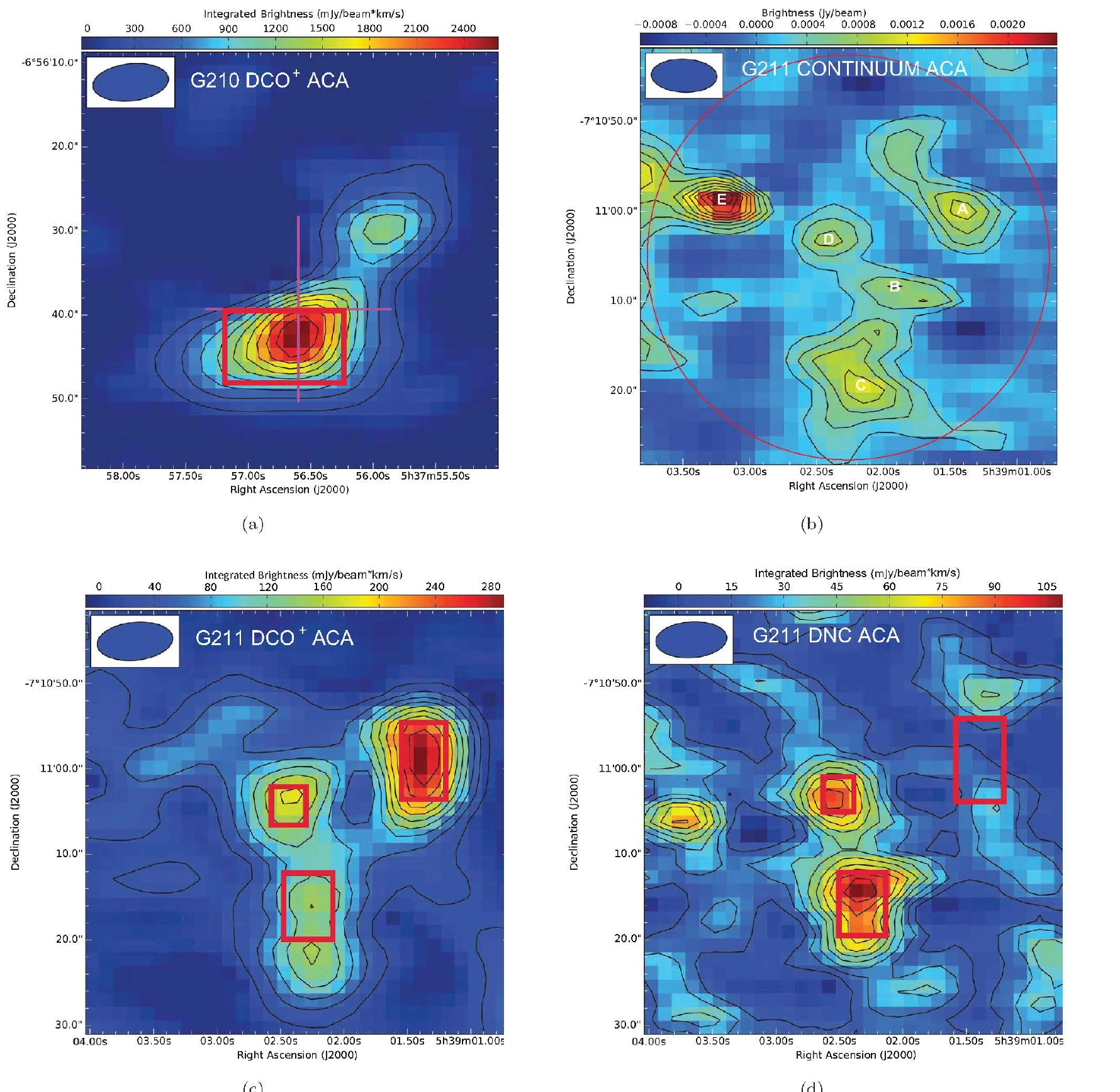}
\caption{Red rectangles depict the averaging areas used for the intensity and the spectrum.
(a) DCO$^+$ map toward G210,
(b) continuum map toward G211,
(c) DCO$^+$ map toward G211, and
(d) DNC map toward G211. }
\end{figure*}

\section{Discussion} \label{sec:discussion}

G210 shows a simple structure, whereas G211 shows complicated structures. In this section, we discuss the properties of each object.

\subsection{G210}

We first discuss the distribution of deuterated molecules in G210 (Figure 4).
Deuterated molecules are less abundant in warmer gas because
the parent molecule for deuterated molecules, H$_2$D$^+$, is easily destroyed by CO evaporated from
the dust in warmer gas \citep{2006AA...460..709F}.
Indeed, gas in the vicinity of a protostar should be warm and as a result
it is reasonable to expect that the distribution of deuterated molecules 
will be depressed toward the protostar.
Nevertheless, Figure 4 shows that DCN, unlike other deuterated molecules, traces warm molecular gas 
near the protostar.
Similarly, \cite{2015A&A...579A..80G} showed that DCN traces warmer regions better than DNC does.
This difference could be the reason for the single-peak DCN distribution.
Non-deuterated molecules do not show the spatial double peaks either (Figure 5).
If optically thin, they likely represent the underlying column density and temperature distributions,
which should increase toward the position of the protostar, in a simple manner.

Figure 9 shows the moment-1, intensity-weighted radial velocity maps of G210.
There is a barely significant velocity difference between the two N$_2$D$^+$ peaks (Figure 9(a)). 
DCO$^+$ $J$ = 3$-$2
and
DNC $J$ = 3$-$2, however, show a velocity gradient from the south-east (blue-shifted) to the north-west (red-shifted)
(Figures 9(b) and (c)).
Because N$_2$D$^+$ is more optically thin, 
it likely traces better
the inner part of the core than DCO$^+$ and DNC do.
Our finding is similar to that by \cite{2018A&A...617A..27P}: N$_2$D$^+$ and N$_2$H$^+$ show smaller 
velocity gradients than the lower density tracers DCO$^+$ and H$^{13}$CO$^+$
in cores in the Taurus dark cloud L1495.
One possible interpretation for the two N$_2$D$^+$ peaks without significant velocity difference is that we detect an edge-on disk
in N$_2$D$^+$ with a radius of $8-10 \arcsec$ or 4000 AU.
With our spatial resolution, however, it is impossible to see the protoplanetary disk ($\sim$ 100 AU).
The DCO$^+$ distribution is larger than the N$_2$D$^+$ distribution.
DCO$^+$ and DNC likely trace the outer part of the core, which, as seen here, is expected to be more turbulent and irregular in distribution and kinematics.
On the other hand, the N$_2$D$^+$ distribution may trace a non-rotating edge-on pseudo-disk.
Figure 10 shows the position-velocity diagram along the strip line passing through
the two N$_2$D$^+$ peaks with a P.A. angle of 51$\degree$.
Looking at the intensity maxima, the velocity difference of the two N$_2$D$^+$ peaks is 
only about 0.17 km s$^{-1}$,
which is close to the spectral resolution (0.16 km s$^{-1}$).
Therefore, the velocity difference is marginal, and we cannot ascertain clearly whether there is a velocity gradient or not.
The radius of 4000 AU is between a large value of $\sim 15,000–30,000$ AU of
the archetypal pseudo-disk associated with the Class 0 protostar L1157 \citep{2007ApJ...670L.131L}
and a small value of  165–192 AU around a proto-brown dwarf \citep{2019MNRAS.486.4114R}.
\cite{1997ApJ...488..317O} investigated the specific angular momentum of the rotating protoplanetary disk and the (collapsing) 
envelope (or pseudo-disk) as well as the molecular cloud core.
The radius of 4000 AU corresponds to a boundary between the two regimes of disks (the specific angular momentum is almost constant) and 
cores (the specific angular momentum decreasing with decreasing radius).
It is possible that G210 has a rotating protoplanetary disk having a size of $\sim$ 100 AU similar to
the disk associated with the low-mass protostar IRAS 04368+2557 in L1527 \citep{2014Natur.507...78S}.
Higher resolution studies are desired to investigate this possibility.

\begin{figure*}
\includegraphics[bb=0 0 600 600, width=20cm]{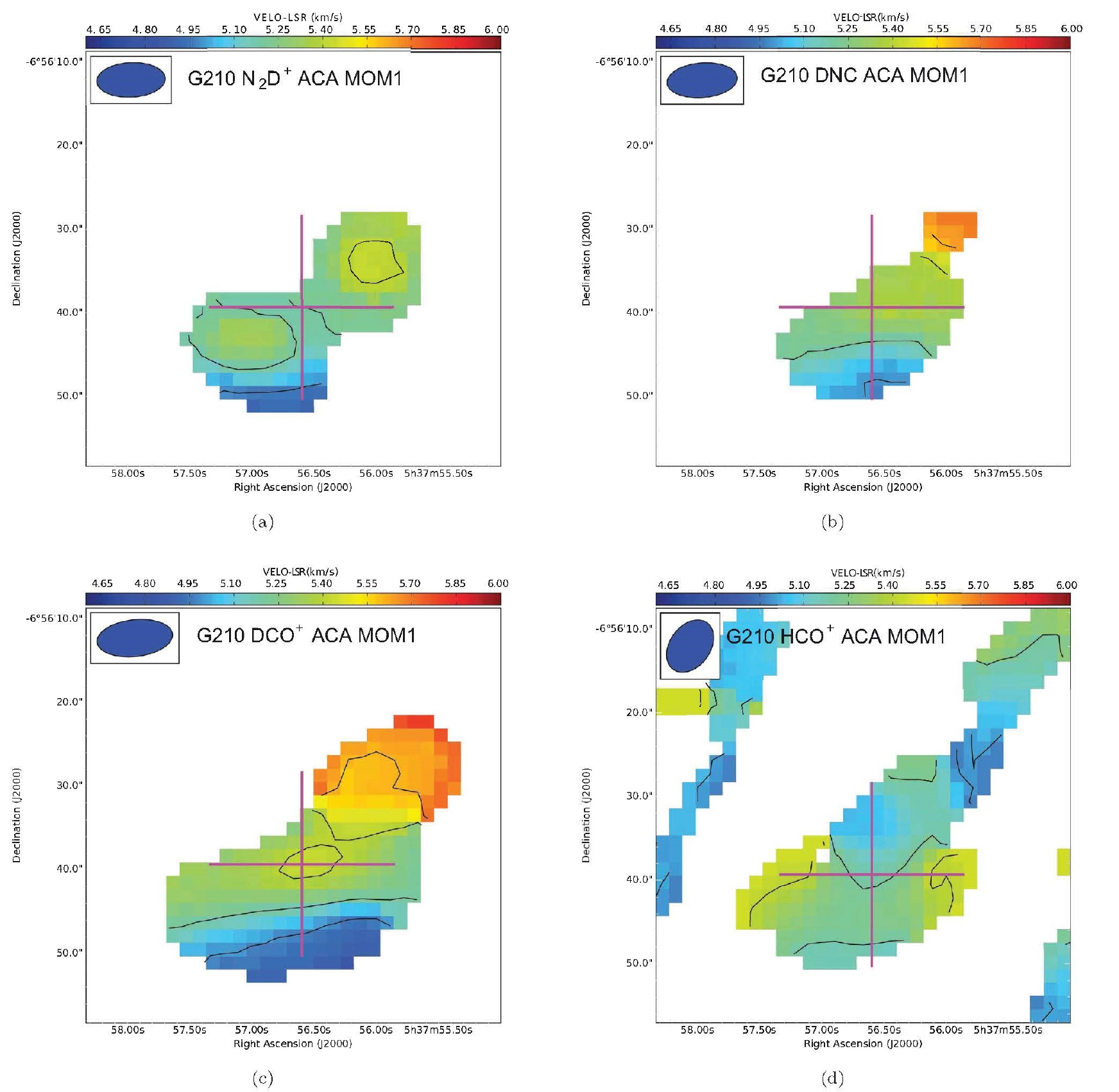}
\caption{The intensity-weighted radial velocity maps of G210 in 
(a) N$_2$D$^+$ $J$ = 3$-$2,
(b) DNC $J$ = 3$-$2,
(c) DCO$^+$ $J$ = 3$-$2,
and
(d) HCO$^+$ $J$ = 3$-$2.
The contour interval is 0.2 km s $^{-1}$.
}
\end{figure*}

\begin{figure*}
\includegraphics[bb=0 0 800 800, width=20cm]{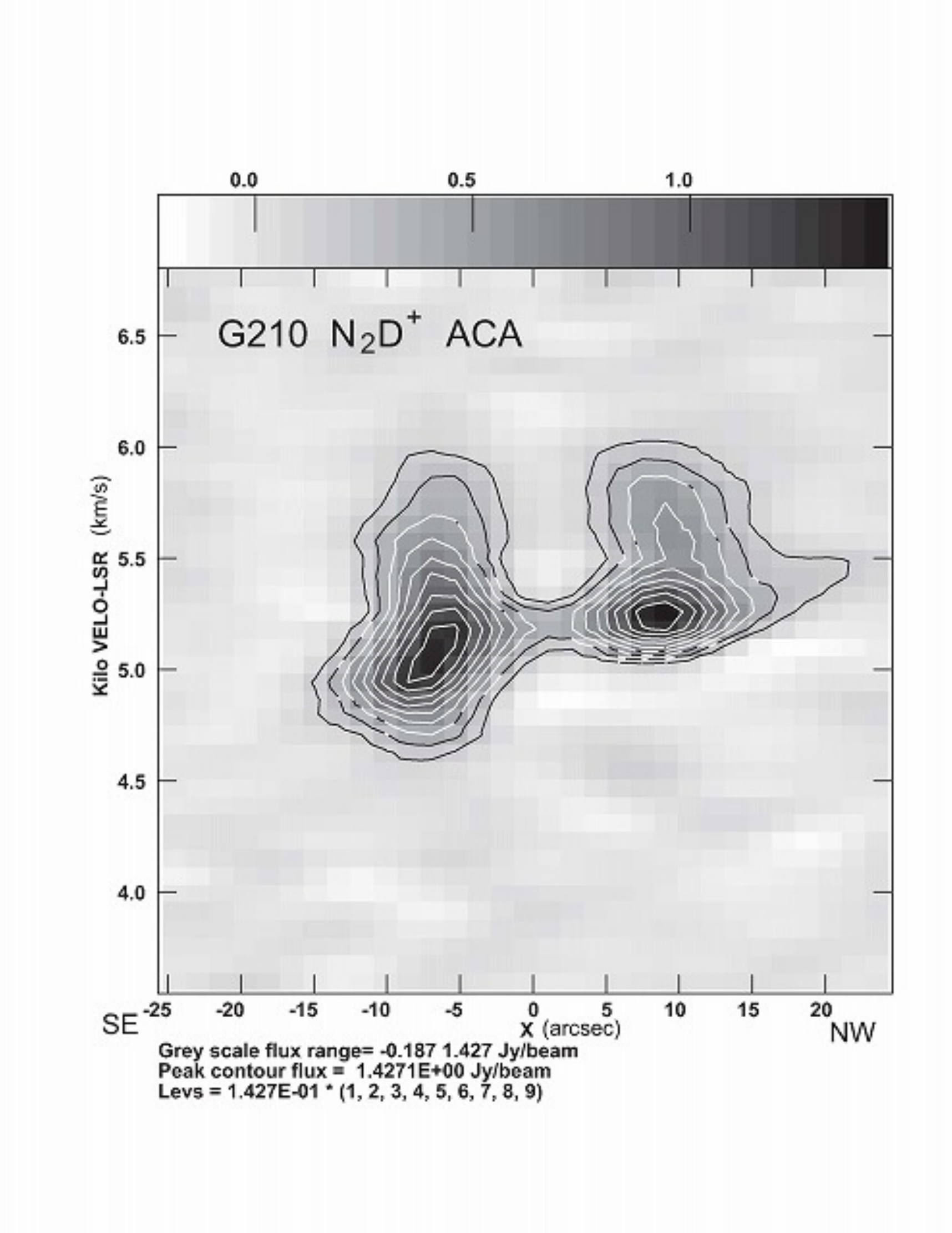}
\caption{
The position-velocity diagram in N$_2$D$^+$ emission along the strip line passing through
two N$_2$D$^+$ peaks of G210 with a P.A. angle of 51$\degree$.
The emission is averaged across the 5$\arcsec$ width perpendicular to the strip line.
The position offset X is defined with respect to the position of the protostar HOPS 157.
}
\end{figure*}

Given that the N$_2$D$^+$ distribution appears to trace a non-rotating (or slowly rotating) edge-on pseudo-disk,
we wonder how the associated molecular outflow looks.
Figure 11 shows the distribution of the high-velocity CO gas.
An edge-on disk is expected to have a symmetric molecular outflow oriented perpendicular to the disk plane;
however, the distribution of the CO lobes is not consistent in this view.
Instead, the red lobe is very compact ($<$ 5$\arcsec$ or $<$ 0.02 pc), while the blue lobe maximum displacement is larger 
(10$\arcsec$ or 0.03 pc).
If we take 18 km s$^{-1}$ and 9 km s$^{-1}$ (the maximum velocity offset from the systemic velocity) 
for the expansion speed of the blue and red lobe, respectively,
the dynamical time scale is as short as 2 $\times$ 10$^3$ yr.
This short duration seems consistent with the interpretation that G210 is immediately after the onset of star formation.
The dynamical time scale here, however, is calculated from the maximum lobe extension and the maximum velocity offset.
This latter value may be quite underestimated for an edge-on system,
while the former may also be underestimated due to missing flux from broader emission.
Indeed, we wonder if the missing flux of CO may affect the appearance of the outflow.
Figure 12 shows the map of outflow from CO ($J$ = 1$-$0) data of the ``Star Formation'' legacy program
with the Nobeyama 45-m telescope \citep{2019PASJ...71S...3N}.
Here the blue lobe is consistently seen, but the red lobe is not.
The widespread red CO emission on the north-western side is associated with the quiescent ambient 
molecular cloud rather than molecular outflows.
Due to this extended CO emission, it is hard to isolate the distribution of the red lobe on this map.
For further comparison, Figure 13 shows the HCO$^+$ outflow map from our ACA observation. 
Relative to CO, it is harder to recognize the outflow configuration, 
but an outflow axis is discernable in the east-west direction.
At any rate, there is no clear evidence that the direction of the outflow lobes is perpendicular to 
the line connecting the two N$_2$D$^+$ peaks
as could be expected if the N$_2$D$^+$ distribution represents a non-rotating (or slowly rotating) edge-on pseudo-disk.
It is thus difficult to explain the N$_2$D$^+$ distribution and the outflow distribution consistently.

Although \cite{2010ApJ...716..893P} and \cite{2016PASJ...68...24T} suggest that there is no
strong correlation among the directions of local magnetic fields, core
elongation, core rotation, and molecular outflows,
we check the orientations of the filament and magnetic field.
The direction of the line connecting the two N$_2$D$^+$ peaks (Figure 4(a), position angle (P.A.) = 51$\degree$) is almost parallel to
that of the larger N$_2$H$^+$ filament containing G210 (Figure 2(a), P.A. = 44$\degree$).
The direction of the global magnetic field for this region is P.A. = 60$\degree$ from the measurement with the {\it Planck} satellite \citep{2018A&A...609L...3S},
almost parallel to the line connecting the two N$_2$D$^+$ peaks and the N$_2$H$^+$ filament.

\begin{figure*}
\includegraphics[bb=0 0 800 800, width=20cm]{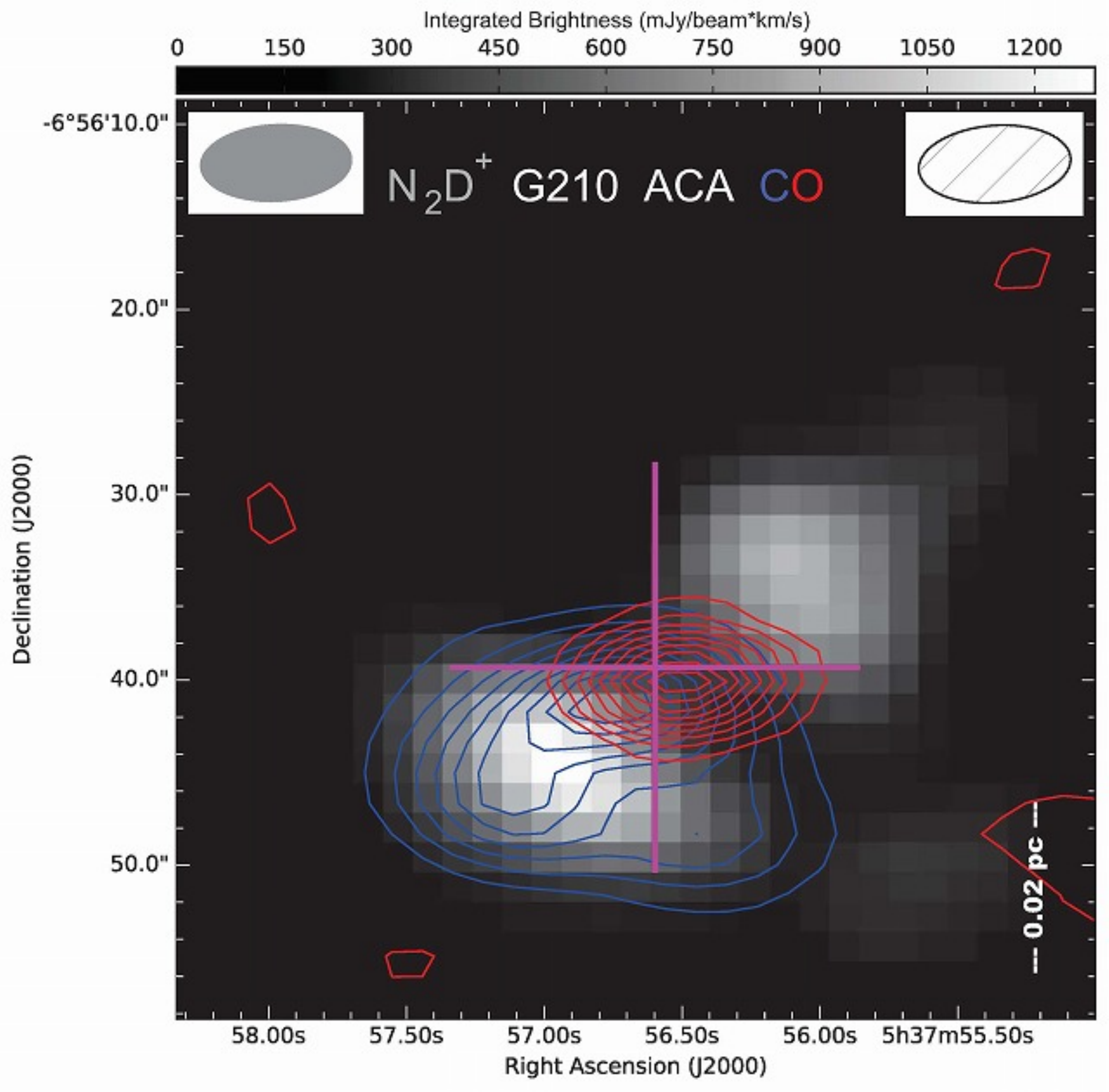}
\caption{
ACA 7m outflow map of CO ($J$ = 2$-$1) toward the star-forming core G210
superimposed on the grey-scale N$_2$D$^+$ map.
The integration ranges are $-$15.0 km s$^{-1}$ to 3.3 km s$^{-1}$ and 10.5 km s$^{-1}$ to 13.0 km s$^{-1}$ for the blue and red lobes, respectively.
The contour interval is  10\% of the maximum intensity ofeach lobe,
where the maximum values are 2.7 Jy beam$^{-1}$ km s$^{-1}$ and: 4.9 Jy beam$^{-1}$ km s$^{-1}$ for blue and red lobes, respectively.
The pink plus symbol shows the location of the continuum peak.
The ovals at the top-left and top-right illustrate the synthesis beam size for the N$_2$D$^+$ and CO data, respectively.
}
\end{figure*}

\begin{figure*}
\includegraphics[bb=0 0 800 800, width=20cm]{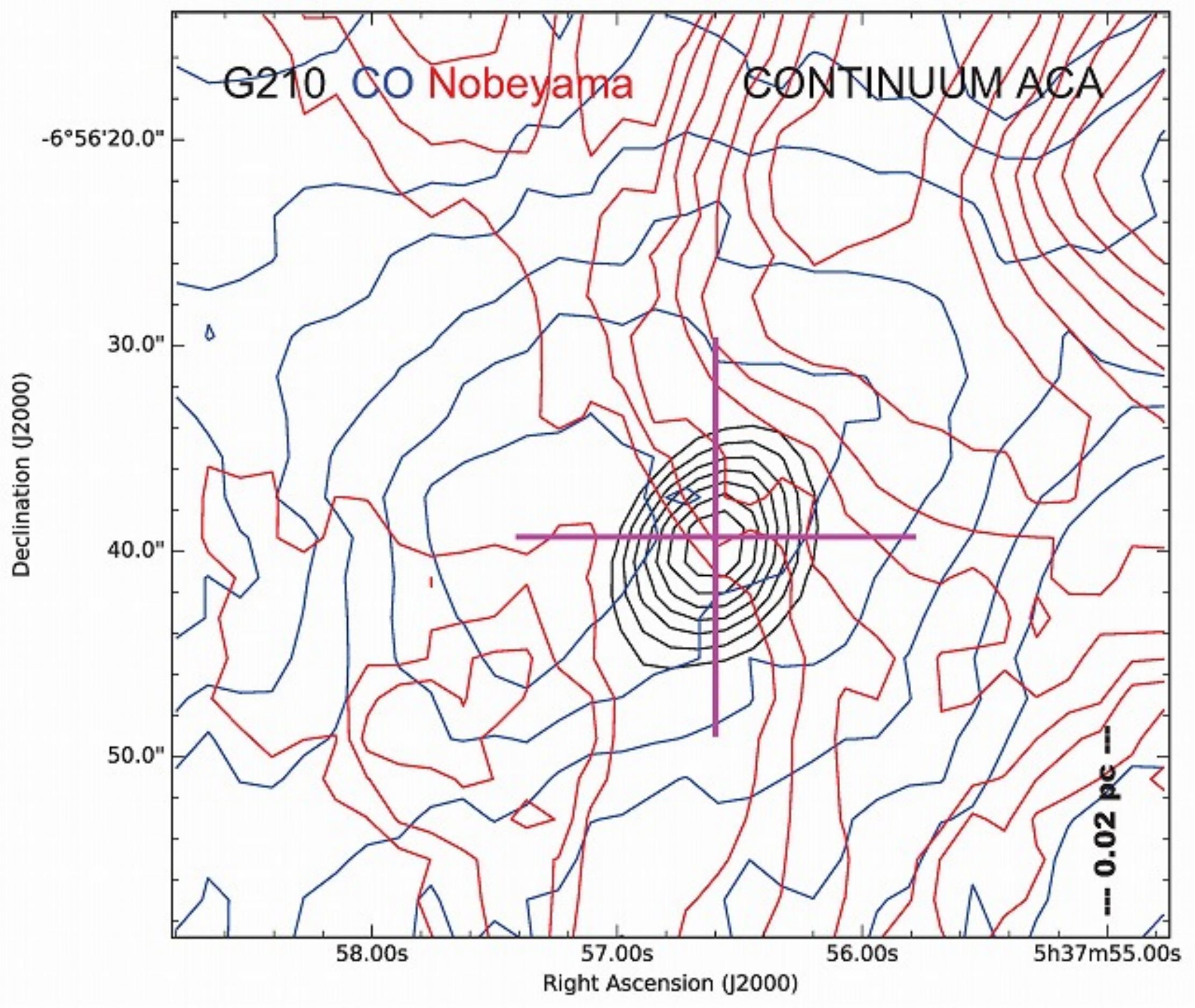}
\caption{
Outflow map of CO ($J$ = 1$-$0) toward the star-forming core G210 from
the data of the Nobeyama 45-m telescope legacy program ``Star Formation''
\citep{2019PASJ...71S...3N}
through the Japanese Virtual Observatory superposed on the continuum map.  
The integration ranges are the same as in Figure 11.
The contour levels are 30, 40, 50, 60, 70, 80, and 90\% of the  maximum intensity of each lobe,
where the maximam values are 2.9 K km s$^{-1}$ and 1.1 K km s$^{-1}$ for the blue and red lobes, respectively.
The plus shows the location of the continuum peak.
The effective spatial resolution is 21$\farcs$7.
}
\end{figure*}

\begin{figure*}
\includegraphics[bb=0 0 800 800, width=20cm]{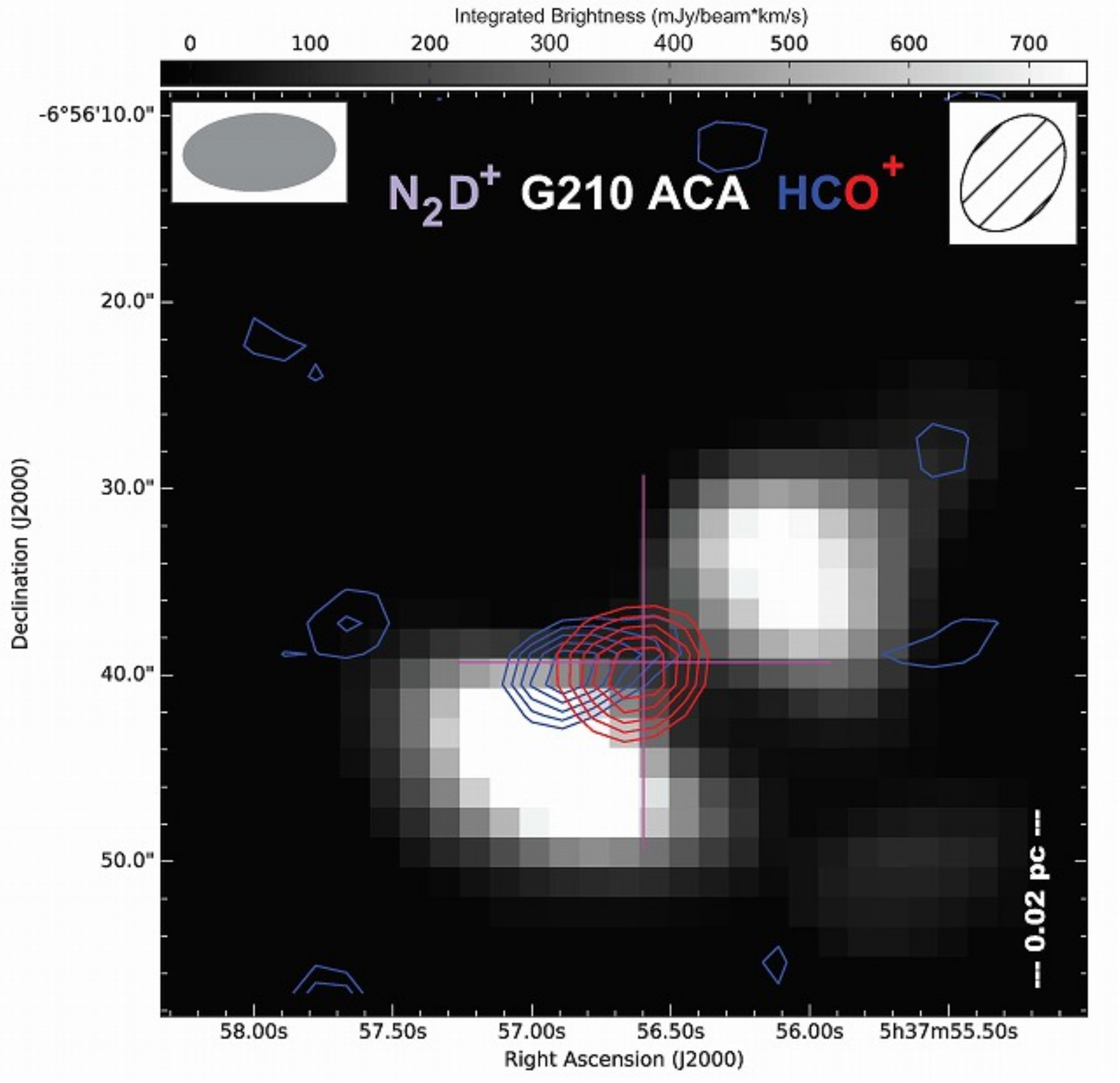}
\caption{
ACA 7m outflow map of HCO$^+$ ($J$ = 3$-$2) toward the star-forming core G210
superimposed on the grey-scale N$_2$D$^+$ map.
The blue and red contours represent the blue and lobes, respectively.
The integration ranges are 2.3$-$4.3 km s$^{-1}$ and 6.0$-$8.0 km s$^{-1}$ for the blue and red lobes, respectively.
The contour levels are 50, 60, 70, 80, and 90\% of the maximum intensity of each lobe,
where the maximam values are 0.75 Jy beam$^{-1}$ km s$^{-1}$ and 0.43 Jy beam$^{-1}$ km s$^{-1}$ for the blue and red lobes, respectively.
The plus shows the location of the continuum peak.
The oval at the top illustrates the synthesis beam size.
}
\end{figure*}

\subsection{G211}

In G211 (Figure 6), we see large differences in the spatial distribution of molecular emission,
which could represent differences in abundances (and optical depths).
To explore this idea, we first compare column densities
calculated through the standard LTE (local thermodynamic equilibrium) method assuming that the line is optically thin.
We assume the gas kinetic temperature to be 16 K from the dust temperature of the associated Planck cold clump.
The H$_2$ volume density estimated from the dust continuum emission is of the order 10$^{6-7}$ cm$^{-3}$,
while the critical densities $n_{cr}$ (H$_2$) of the observed $J$ = 3$-$2 lines are of order 10$^{7-8}$ cm$^{-3}$.
To obtain estimates of the excitation temperature, we ran
the RADEX software\footnote{http://var.sron.nl/radex/radex.php} \citep{2007A&A...468..627V}.
From test values of $N$(H$_2$) = $3\times10^6$ cm$^{-3}$, $T_{kin}$ = 20 K, 
$\Delta v$ (DCO$^+$ = 0.5 km s$^{-1}$, $N$(DCO$^+$) = $(1-10)\times10^{12}$ cm$^{-2}$,
and
$T_{R}$ (DCO$^+$ $J$ = 3$-$2) = 0.3$-$1 K,
we obtain $T_{ex}$ = 15 K (almost fully thermalized) and assume this value for all molecular lines observed with the ACA.
Table 8 summarizes the resulting column densities and abundances.
For a comparison, we also include G210.
We see large variation in the column density ratios among sub-cores in G211 and between G210 and G211.
The HCO$^+$ profiles show hints of (self) absorption (double peaks, Figures 14, and 16), and the line optical depth should be very large.
Therefore, we exclude this molecule from the column density estimation.
Regarding the relative evolutionary stage of the sub-cores, 
the non-detection of N$_2$D$^+$ emission in G211A suggests this sub-core is young, while
relatively bright N$_2$D$^+$ emission in G211C suggests it is a more evolved starless core.
We judge that, among G211A, C, and D,
G211A is the youngest, and G211C is the most evolved starless core,
from the relative intensity of the N$_2$D$^+$ emission compared with the DCO$^+$ and HCO$^+$ emission.
Note also that G211E is not prominent in molecular lines.
It is likely that G211E shows local maximum of the core column density, but
its density is not high enough to excite the $J$ = 3 level of the observed molecules.

\setcounter{table}{7}
\begin{table*}[h!]
\renewcommand{\thetable}{\arabic{table}}
\caption{Column Density and Abundance from ACA}
\center
\begin{tabular}{lllllll}
\hline 
\hline
		      &	$N$(N$_2$D$^+$)	&	$N$(DNC)	&	$N$(DCO$^+$)	&	$X$(N$_2$D$^+$)	&	$X$(DNC)	&	$X$(DCO$^+$)	\\
		      &		                          &		             &		                   &		                         &		             &	                    	\\
Core		&	    cm$^{-2}$	             &	cm$^{-2}$	&	cm$^{-2}$      	&	                           	&	             	&	                          \\ 
\hline																								 		
G210		&	1.8E+12              	&	3.5E+11	     &	1.0E+12	             &	6.2E-12                   	&	1.2E-12     	&	        3.5E-12	\\
																										
G211	A	&	. . .	                         &	. . .	            &	3.4E+11         	&	. .                            .	&	.            . .	&	1.0E-11        	\\
																										
G211	C	&	2.0E+11	                   &	1.1E+11    	&	1.4E+11         	&	1.5E-11               	&	8.7E-12   	&	1.1E-11	            \\
G211	D	&	2.6E+11               	&	1.1E+11   	&	1.9E+11         	&	3.0E-12              	&	1.2E-12	     &	2.1E-12             	\\
\hline
\end{tabular}
\tablecomments{The relative errors of the column density and abundance are 30\%.}
\end{table*}

G211A, C, and D have different velocities (3.48 km s$^{-1}$, 3.64 km s$^{-1}$, and 3.19 km s$^{-1}$, respectively, in DCO$^+$).
Sub-cores G211A, C, and D have FWHM linewidths of 0.22$-$0.29 km s$^{-1}$ (Table 7).
Because the FWHM thermal linewidth of DCO$^+$ for 20 K is 0.17 km s$^{-1}$,
sub-cores do not appear purely thermal.
G211 is very clumpy and its substructures have different chemical compositions and radial velocities,
suggesting that the G211 core is not dynamically and chemically settled, and consists of velocity-coherent sub-cores.

\subsection{Fragmentation, Virial Mass, and Motion}

\cite{2018ApJ...856..147O} found a hint of thermal Jeans fragmentation toward 
the starless core TUKH122 in the same cloud
as G210 and G211, Orion A.
For example, the separation of its condensations is $\sim$ 0.035 pc, consistent with the thermal Jeans length at a density of 
$4.4\times10^5$ cm$^{-3}$ and a temperature of 12 K.
For comparison, the dust continuum map of G211 (Figure 3 (b)) shows separations among sub-cores of 
about 10$\arcsec$ or 0.02 pc,
similar to the value of \cite{2018ApJ...856..147O}.
hence, our data support an idea of thermal Jeans fragmentation in the prestellar phase.
In the Taurus molecular cloud, which is the archetypal nearby dark cloud,
we observe examples of fragmentation of filaments.
For example, the Miz2, Miz7, Miz8b, and Miz8 cores \citep{2004ApJ...606..333T} are spaced at equal intervals of $\sim$ 1.3 arcmin or 0.07 pc 
along their host filament.
If we assume different densities of $1\times10^5$ \citep{2002ApJ...575..950O}, $3\times10^5$ \citep{2014ApJ...789...83T}, and  $1\times10^6$ cm$^{-3}$ and a common temperature of 10 K
for Taurus cloud cores, TUKH122-n in Orion, and  G211 (judging from those for G211A and G211C), respectively, the spatial spacing can be interpreted in terms of
thermal Jeans fragmentation.
\cite{2019ApJ...886..102S} also reported thermal Jeans fragmentation in IRDC clumps.
Moreover, we note that the observed FWHM linewidth of N$_2$D$^+$ is 0.32-0.34 km s$^{-1}$ for G211, and 
the corresponding effective sound speed (line-of-sight velocity dispersion $\sigma$ = FWHM linewidth/$\sqrt{ 8 ln 2}$) corresponding to 2.33 u
(the mean molecular weight per particle in terms of the atomic mass unit) \citep{1992ApJ...384..523F,2000ApJS..131..465A}
is 0.23 km s$^{-1}$, which is very close to the sound speed for 10 K (0.19 km s$^{-1}$).
It seems that turbulence has been almost completely dissipated in these sub-cores.

Next, we discuss the virial masses of G210 and G211.  The virial mass $M_{vir}$
is calculated from the radius $R$ (HWHM = FWHM/2)
deconvolved for the synthesized beam size of the dust continuum map
and the FWHM linewidth in DCO$^+$,
assuming the constant density sphere \citep{1988ApJ...333..821M}.
We adopt DCO$^+$, because its line emission is detected from both cores and delineates
the core relatively well.
The virial parameter $\alpha_{vir}$ is defined as  $\alpha_{vir} = M_{vir}/M_{dust}$.

G211 is divided into several sub-cores (0.01$-$0.07 \nom{M})  having virial parameters of 1.5$-$2.4.
In high-mass star-forming regions, small virial parameters have been reported \citep{2013ApJ...779..185K};
however, this is not the case for G211.
The evolved starless core G211C has a larger virial parameter ($\alpha_{vir}$ = 2.0)
than the younger starless core G211A ($\alpha_{vir}$ = 1.5).
G210 is a solar-mass object, and its virial parameter is 0.5.
However, the sample number is too small for a statistical study of the virial parameter of the sub-core, and further studies are desired. 
In the larger scale, CS ($J$ = 1$-$0) core TUKH101 \citep{1993ApJ...404..643T} corresponding to G211 
has an LTE mass $M_{LTE}$ of 27 \nom{M} and a virial mass $M_{vir}$ of 30 \nom{M}, and the virial parameter $\alpha_{vir} = M_{vir}/M_{LTE}$
is close to unity.
\cite{2017ApJ...834..193K} studied high-mass starless cores in IRDCs with the ALMA in the continuum and lines including N$_2$D$^+$, 
and concluded that the six ``best'' N$_2$D$^+$ cores are consistent with virial equilibrium of pressure confined cores.  
The virial parameter ranges from 0.2 to 0.75.
We wonder if there are environmental differences among nearby dark clouds, general GMCs (regions outside IRDCs), and IRDCs, because
the external pressure is likely to differ \citep{1993ApJ...404..643T,2000ApJS..131..465A,2009ApJ...699.1092H}.

\setcounter{table}{8}
\begin{table*}[h!]
\renewcommand{\thetable}{\arabic{table}}
\caption{Mass and virial mass from ACA}
\center
\begin{tabular}{lllll}
\hline 
\hline 										
core		&	M$_{dust}$	&	$\Delta v$(DCO$^+$)	&	$M_{vir}$	&	$\alpha_{vir}$	\\
		&	$\nom{M}$	&	km s$^{-1}$	&	$\nom{M}$	&		\\
\hline										
G210		&	0.72 	&	0.55 	&	0.39 	&	0.5 	\\
G211	(all)	&	0.15 	&	. . .	&	. . .	&	. . .	\\
G211	A	&	0.019 	&	0.22 	&	0.030 	&	1.5 	\\
										
G211	C	&	0.07 	&	0.28 	&	0.15 	&	2.1 	\\
G211	D	&	0.009 	&	0.29 	&	0.022 	&	2.4 	\\
\hline
\end{tabular}
\tablecomments{The dust continuum radius $R_{dust}$ is corrected for the synthesized beam size.  The radius error is about 20\%. The errors  of the mass and virial parameter are about 30\%. The linewidth error is about 10\%.}
\end{table*}

Finally, we explore whether there is any hints of collapse or expansion motions associated with G210 and G211 cores.
Figures 14 to 17 represent 
the line profiles averaged over the specific area for G210 and G211 depicted in Figure 8.
Occasionally, we see absorption due to missing flux for extended emission.
For HCO$^+$, the absorption is severe.
We see the wing emission in HCO$^+$ toward G210.
For the continuum emission,
41\% and 21\% of the SCUBA-2 flux densities were recovered with the ACA 7 m Array for G210 and G211, respectively.
Deuterated molecules trace high-density ($\ga 10^5$ cm$^{-3}$)  gas, and their intensity distribution is rather compact.
Therefore, the effect of missing flux  should be small.

The DCO$^+$ line profile toward the starless sub-core G211D (Fig. 17(c)) shows a signature of the inverse P Cygni profile.
The absorption feature ($v_{LSR}$ = 3.695$\pm$0.035 km s$^{-1}$ through Gaussian fitting) is red-shifted with respect to the radial velocity of the core ($v_{LSR}$ = 3.186$\pm$0.006 km s$^{-1}$) by 0.509$\pm$0.036 km s$^{-1}$. 
It is possible that the inverse P Cygni profile represents infall motions,
which means that this sub-core is still at the stage of mass accretion.
The dust continuum radius $R_{dust}$ = 0.0012 pc and the DCO$^+$ linewidth $\Delta v$ (DCO$^+$) = 0.29 km s$^{-1}$ provide a time scale of 4$\times$10$^3$ yr.
This value and the dust continuum mass $M_{dust}$ = 0.009 \nom{M} lead to a mass accretion rate of 2$\times10^{-6} \nom{M}$ yr$^{-1}$.
\cite{2018ApJ...861...14C} also reported infall motions toward a prestellar core in an IRDC.
The dip velocity of $\sim$ 3.8 km s$^{-1}$ in the DCO$^+$ profile is appreciably different from
that in HCO$^+$ of $\sim$ 3.0 km s$^{-1}$, which is probably due to the missing flux.
On the other hand, the velocities in the N$_2$D$^+$ and DNC emission are $\sim$ 3.0 km s$^{-1}$.
It is possible that the inverse P Cygni profile cannot be explained in terms of missing flux, but represents the infall motion.

\bigskip
\bigskip

\begin{figure*}
\includegraphics[bb=0 0 600 600, width=20cm]{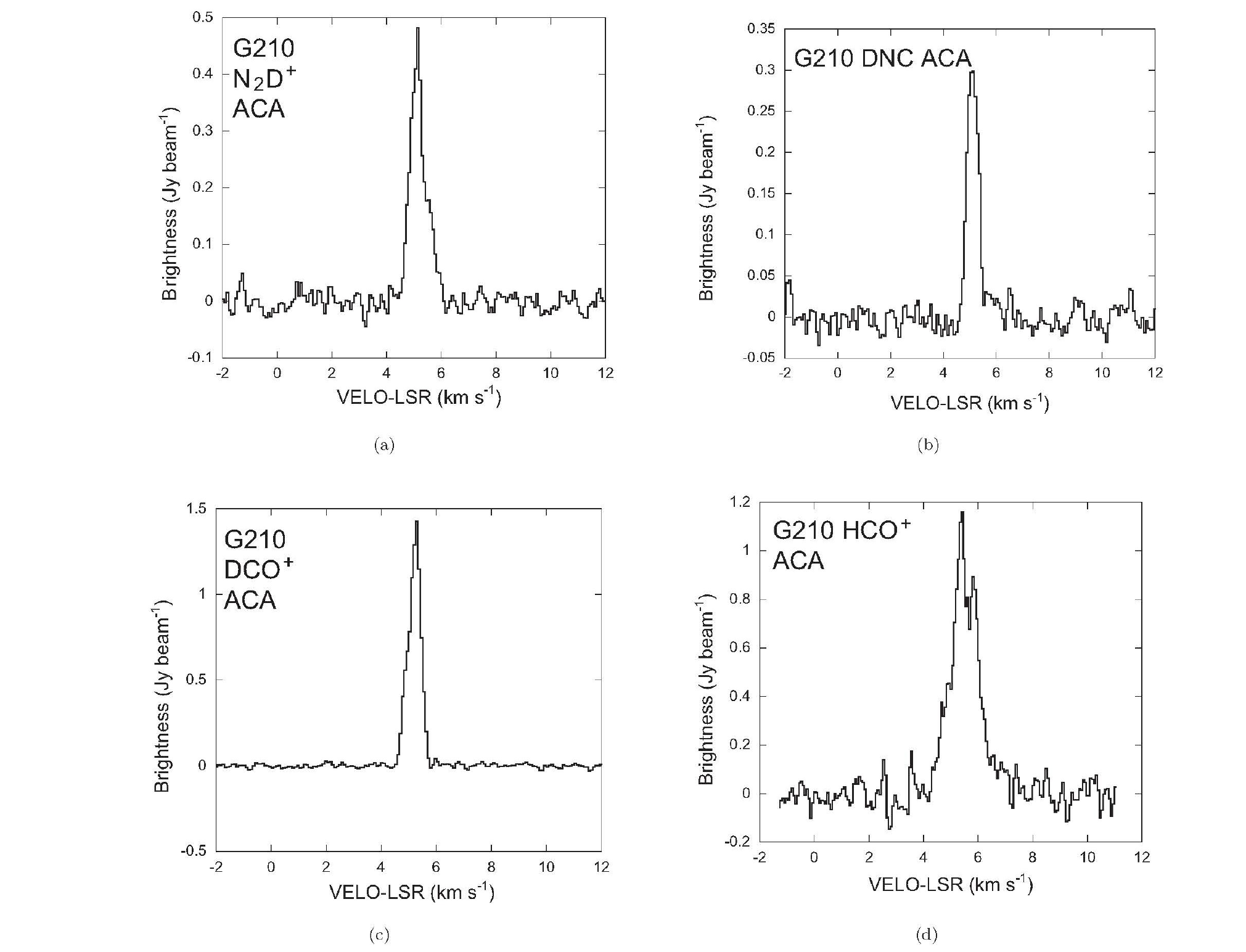}
 \caption{Line profiles of G210 in 
(a) N$_2$D$^+$ $J$ = 3$-$2.
(b) DNC $J$ = 3$-$2,
(c) DCO$^+$ $J$ = 3$-$2,
and
(d) HCO$^+$ $J$ = 3$-$2.
}
\end{figure*}

\begin{figure*}
\includegraphics[bb=0 0 600 600, width=20cm]{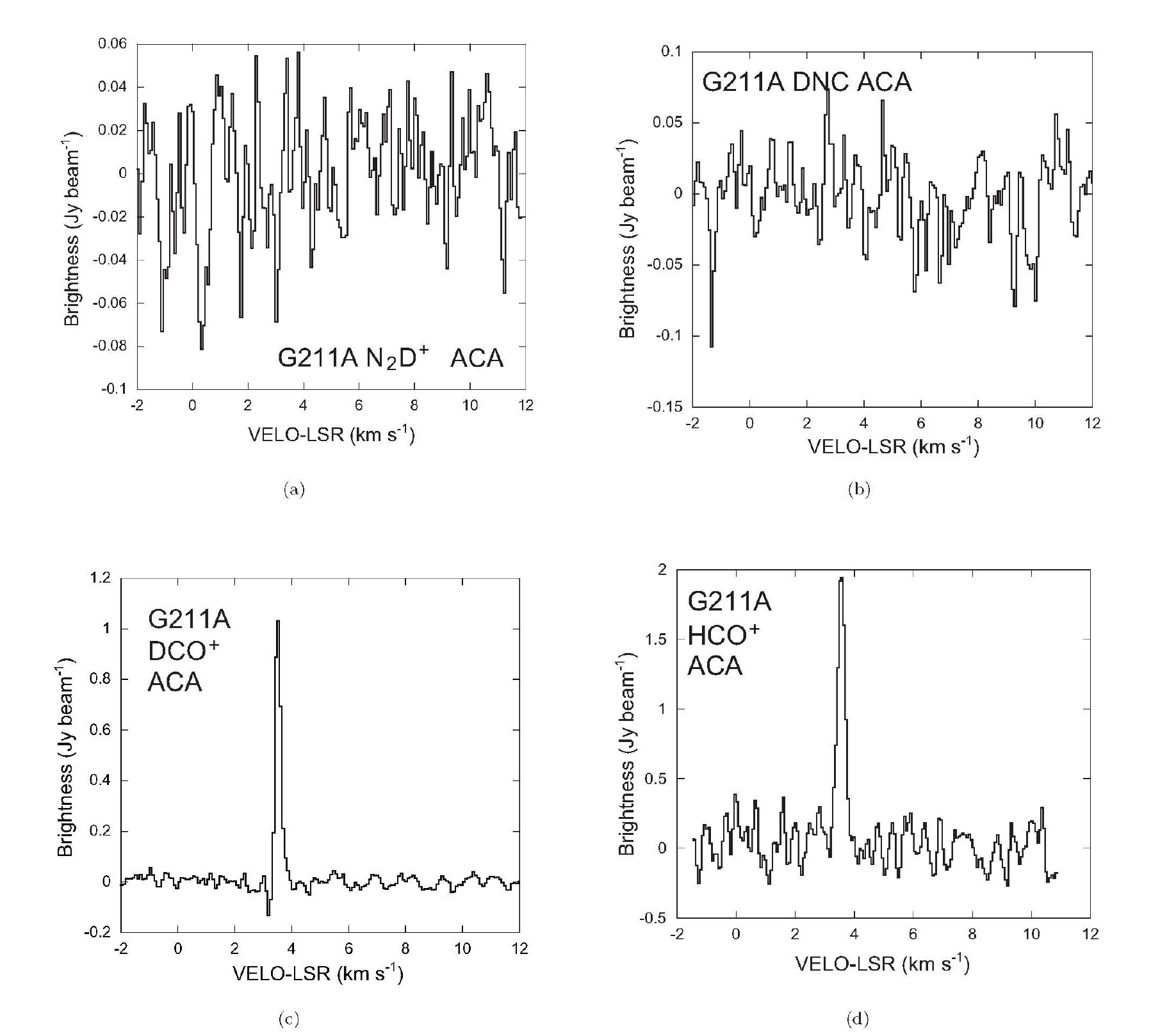}
\caption{Line profiles of G211A in 
(a) N$_2$D$^+$ $J$ = 3$-$2.
(b) DNC $J$ = 3$-$2,
(c) DCO$^+$ $J$ = 3$-$2,
and
(d) HCO$^+$ $J$ = 3$-$2.
}
\end{figure*}

\begin{figure*}
\includegraphics[bb=0 0 600 600, width=20cm]{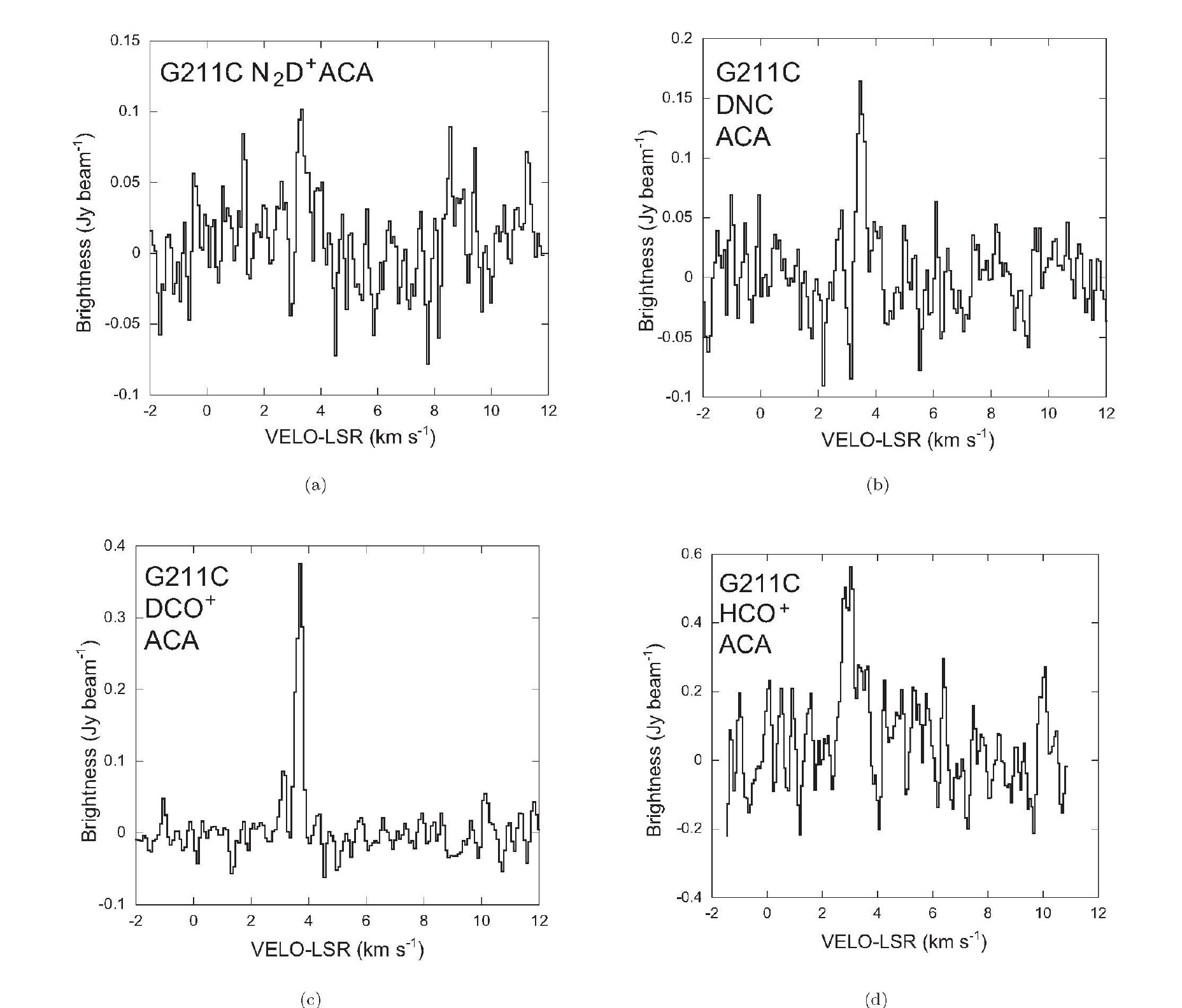}
\caption{Line profiles of G211C in 
(a) N$_2$D$^+$ $J$ = 3$-$2.
(b) DNC $J$ = 3$-$2,
(c) DCO$^+$ $J$ = 3$-$2,
and
(d) HCO$^+$ $J$ = 3$-$2.
}
\end{figure*}

\begin{figure*}
\includegraphics[bb=0 0 600 600, width=20cm]{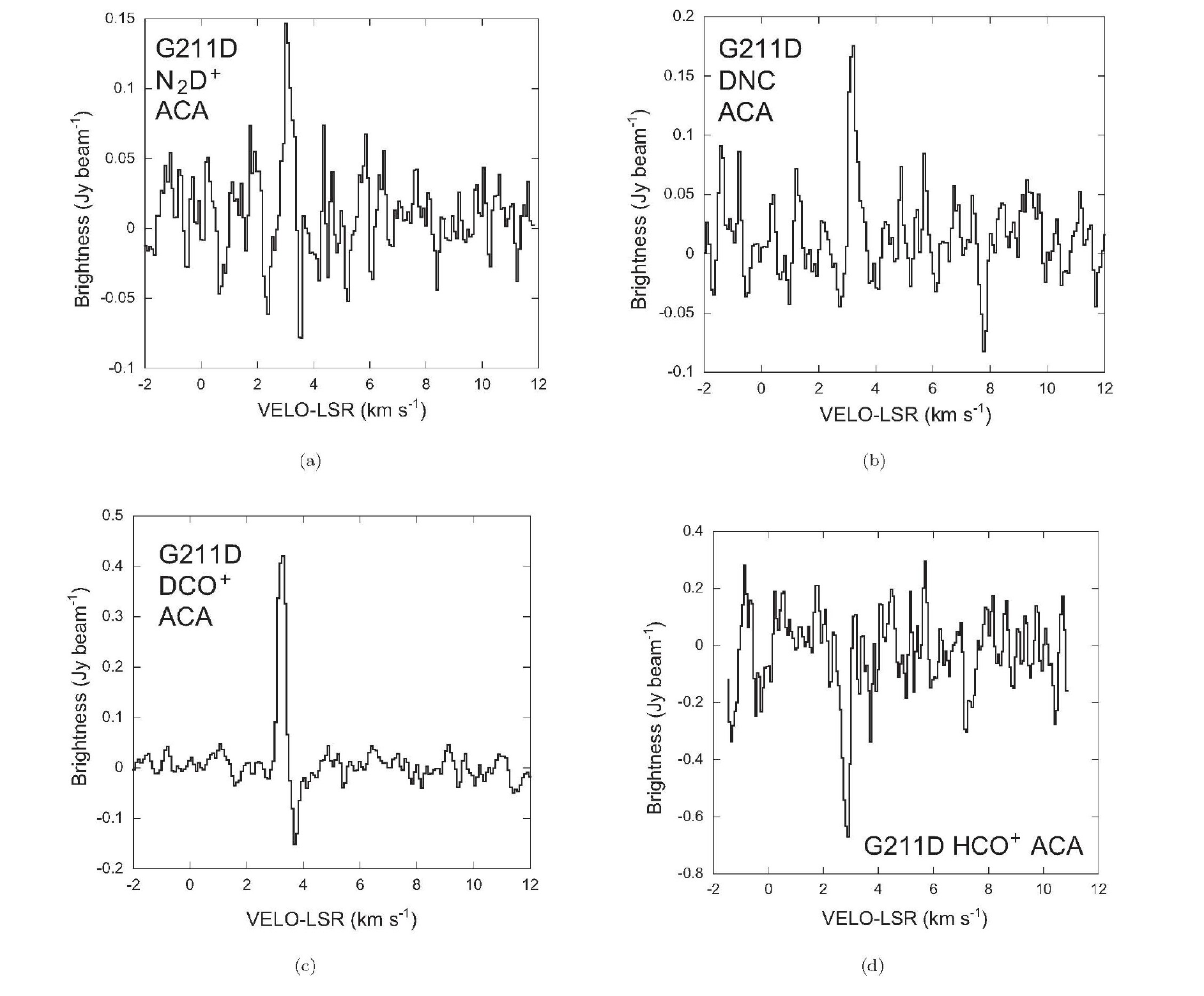}
\caption{Line profiles of G211D in 
(a) N$_2$D$^+$ $J$ = 3$-$2.
(b) DNC $J$ = 3$-$2,
(c) DCO$^+$ $J$ = 3$-$2,
and
(d) HCO$^+$ $J$ = 3$-$2.
}
\end{figure*}


\section{Summary}

We mapped two molecular cloud cores, G210 and G211, in the Orion A cloud 
with the ALMA 7-m Array and with the Nobeyama 45-m radio telescope.  
They are bright N$_2$D$^+$ cores selected from single-pointing observations with the Nobeyama 45-m radio telescope,
and are thought to be relatively close to the onset of star formation.
G210 is a star-forming core and G211 is starless.
These cores
show linewidths of 0.41 km s$^{-1}$ and 0.45 km s$^{-1}$ in N$_2$D$^+$ 
in single-dish observations with the Nobeyama 45-m telescope.
Both cores were detected with ALMA ACA 7-m Array  in the continuum and molecular lines at Band 6.
The star-forming core G210 shows an interesting spatial feature of N$_2$D$^+$: 
two N$_2$D$^+$ peaks with similar intensities and radial velocities
are located symmetrically around the dust continuum peak.
One interpretation is that the two N$_2$D$^+$ peaks
represent an edge-on pseudo-disk, which has little-to-no detectable rotation.
The starless core G211 shows very clumpy substructure and contains multiple sub-cores,
which show appreciable chemical differences.
The sub-cores in G211 have internal motions that are almost purely thermal.
The starless sub-core G211D in particular shows a hint of the inverse P Cygni profile, suggesting infall motions.

\acknowledgments

This paper makes use of the following ALMA data: ADS/JAO.ALMA\#2016.2.00058.S. ALMA is a partnership of ESO (representing its member states), NSF (USA) and NINS (Japan), together with NRC (Canada), MOST and ASIAA (Taiwan), and KASI (Republic of Korea), in cooperation with the Republic of Chile. The Joint ALMA Observatory is operated by ESO, AUI/NRAO and NAOJ.
Data analysis was carried out on the open use data analysis computer system at the Astronomy Data Center, 
ADC, of the National Astronomical Observatory of Japan.
Some data were retrieved from the JVO portal\footnote{http://jvo.nao.ac.jp/portal/} operated by ADC/NAOJ.
K.T. thanks Satoshi Yamamoto and Nami Sakai for discussion.
P.S. was partially supported by a Grant-in-Aid for Scientific Research (KAKENHI Number 18H01259) 
of Japan Society for the Promotion of Science (JSPS).

%

\vspace{5mm}
\facilities{ALMA,No:45m}


\software{AIPS \citep{1996ASPC..101...37V},  
          CASA \citep[v4.7.2 and v5.4.0;][]{2007ASPC..376..127M},
          NewStar,
          NoStar,
	    RADEX \citep{2007A&A...468..627V}
}

\end{document}